\documentclass[preprint]{aastex}
\slugcomment{submitted to the Astronomical Journal}
\shortauthors{Crosthwaite & Turner}
\shorttitle{CO and Neutral Gas in NGC~6946}

\newcommand{\rah}[1]{\mbox{$\alpha\,=\,#1^{h}$}}
\newcommand{\ram}[1]{\mbox{$#1^{m}$}}
\newcommand{\ras}[1]{\mbox{$#1^{s}$ }}
\newcommand{\decd}[1]{\mbox{$\delta\,=\,#1^\circ$}}
\newcommand{\decm}[1]{\mbox{$#1^{'}$}}
\newcommand{\decs}[1]{\mbox{$#1^{''}$ }}

\newcommand{\du}{\mbox{cm$^{-3}$}}

\newcommand{\cdu}{\mbox{cm$^{-2}$}}

\newcommand{\Nmh}{\mbox{$N_{H_{2}}$}}

\newcommand{\sdu}{\mbox{M$_{\sun}$ pc$^{-2}$}}

\newcommand{\sdmh}{\mbox{$\Sigma_{H_{2}}$}}
\newcommand{\sdhi}{\mbox{$\Sigma_{HI}$}}
\newcommand{\sdg}{\mbox{$\Sigma_{gas}$}}

\newcommand{\sdgeqn}{\mbox{$\sdg\! = 1.36\:(\sdhi\! + \sdmh\!)$}}

\newcommand{\sfrfir}{\mbox{SFR$_{fir}$}}
\newcommand{\sfrntc}{\mbox{SFR$_{ntc}$}}
\newcommand{\sfru}{\mbox{M$_{\sun}$ yr$^{-1}$}}
\newcommand{\sdsfr}{\mbox{$\Sigma_{SFR}$}}
\newcommand{\sdsfrfir}{\mbox{$\Sigma_{SFR_{fir}}$}}
\newcommand{\sdsfrntc}{\mbox{$\Sigma_{SFR_{ntc}}$}}
\newcommand{\sdsfrkpcu}{\mbox{M$_{\sun}$ yr$^{-1}$ kpc$^{-2}$}}
\newcommand{\sdsfrpcu}{\mbox{M$_{\sun}$ yr$^{-1}$ pc$^{-2}$}}
\newcommand{\Tsfe}{\mbox{$\tau_{SFE}$}}
\newcommand{\gasSF}{\mbox{Gas$_{tracer}$/SF$_{tracer}$}}

\newcommand{\ha}{\mbox{H${\alpha}$}}
\newcommand{\mh}{\mbox{H$_{2}$}}

\newcommand{\Mmh}{\mbox{M$_{H_{2}}$}}
\newcommand{\Mhi}{\mbox{M$_{HI}$}}

\newcommand{\Mo}{\mbox{M$_{\sun}$}}

\newcommand{\vu}{\mbox{km s$^{-1}$}}

\newcommand{\Trs}{\mbox{$T_{r}^{*}$}}
\newcommand{\Tmb}{\mbox{$T_{mb}$}}

\newcommand{\Tex}{\mbox{$T_{ex}$}}

\newcommand{\Tk}{\mbox{$T_{k}$}}

\newcommand{\Toz}{\mbox{$T_{10}$}}
\newcommand{\Tto}{\mbox{$T_{21}$}}
\newcommand{\TmbTrseqn}{\mbox{$T_{mb} = T_{r}^{*}/\eta^{*}_{m}$}}
\newcommand{\Tmbeqn}{$$ \Tmb =\; \fa \; \frac{c^{2}}{2 k \nu^{2}} \;
                     ( \Buex - \Bucmb) \; ( 1 - e^{-\tau_{\nu}})\:, $$}

\newcommand{\nmb}{\mbox{$\eta^{*}_{m}$}}

\newcommand{\Rco}{\mbox{$r_{12}$}}

\newcommand{\iu}{\mbox{K km s$^{-1}$}}
\newcommand{\Ico}{\mbox{$I_{CO}$}}
\newcommand{\Ioz}{\mbox{$I_{10}$}}
\newcommand{\Ito}{\mbox{$I_{21}$}}

\newcommand{\Lfir}{\mbox{$L_{fir}$}}
\newcommand{\Lntc}{\mbox{$L_{ntc}$}}
\newcommand{\Ihi}{\mbox{$I_{HI}$}}

\newcommand{\Iozeqn}{\mbox{$\Ioz\! = \int \Toz\: dv$}}

\newcommand{\su}{\mbox{Jy beam$^{-1}$ km s$^{-1}$}}
\newcommand{\sut}{\mbox{Jy km s$^{-1}$}}
\newcommand{\jyb}{\mbox{Jy beam$^{-1}$}}
\newcommand{\mjyb}{\mbox{mJy beam$^{-1}$}}
\newcommand{\Mjys}{\mbox{MJy sr$^{-1}$}}
\newcommand{\jys}{\mbox{Jy sr$^{-1}$}}
\newcommand{\Wms}{\mbox{W m$^{-2}$ sr$^{-1}$ }}
\newcommand{\Whz}{\mbox{W Hz$^{-1}$}}
\newcommand{\Watt}{\mbox{W}}

\newcommand{\ftocm}{\mbox{S$_{21 cm}$}}

\newcommand{\pu}{\mbox{cm$^{-3}$ K}}
\newcommand{\Pism}{\mbox{$P_{ism}$}}
\newcommand{\Pmin}{\mbox{$P_{min}$}}
\newcommand{\fa}{\mbox{f$_{a}$}}

\newcommand{\tu}{\mbox{$\tau_{\nu}$}}

\newcommand{\Bu}{\mbox{$B_{\nu}$}}
\newcommand{\Buex}{\mbox{$B_{\nu}(T_{ex})$}}
\newcommand{\Bucmb}{\mbox{$B_{\nu}(T_{cmb})$}}
\newcommand{\Xco}{\mbox{$X_{CO}$}}
\newcommand{\fir}{\mbox{$\mu$m}}

\newcommand{\Xcoeqn}{\mbox{$\Xco = \Nmh/\Ico = 2 \times 10^{20}\;
                                   \cdu\; (\iu)^{-1}$ }}
\newcommand{\Xcoval}[1]{\mbox{$\Xco = #1\: \times 10^{20}\;
                                      \cdu\; (\iu)^{-1}$ }}

\newcommand{\Dcor}{\mbox{(D/6 Mpc)}}
\newcommand{\Dcorsqr}{\mbox{(D/6 Mpc)$^2$}}

\newcommand{\fcs}{\setlength{\baselineskip}{0.6cm}}

\begin{document}

\title{ CO(1-0), CO(2-1) and Neutral Gas in NGC~6946: \\
        Molecular Gas in a Late-Type, Gas Rich, Spiral Galaxy }
\author{ Lucian P. Crosthwaite\altaffilmark{1} }
\affil{ Northrop Grumman, Unmanned Systems, San Diego, CA 92150 }
\and
\author{ Jean L. Turner\altaffilmark{2} } 
\affil{ Division of Astronomy and Astrophysics, UCLA, Los Angeles, CA 90095 }

\altaffiltext{1}{ lpcrosthwaite@cox.net }
\altaffiltext{2}{ turner@astro.ucla.edu }

\begin{abstract}

We present ``On The Fly" maps of the CO(1-0) and CO(2-1) emission 
covering a 10\arcmin~$\times$~10\arcmin\ region of the NGC~6946. 
Using our CO maps and archival VLA HI observations we create
a total gas surface density map, \sdg, for NGC~6946. 
The predominantly molecular inner gas disk transitions smoothly
into an atomic outer gas disk, with equivalent atomic and molecular 
gas surface densities at R = 3.5\arcmin\ (6 kpc).
We estimate that the total \mh\ mass is $3\times10^{9}$ \Mo, 
roughly 1/3 of the interstellar hydrogen gas mass, 
and about 2\% of the dynamical mass of the 
galaxy at our assumed distance of 6 Mpc. 
The value of the CO(2-1)/CO(1-0) line ratio ranges from 0.35 to 2; 
50\% of the map is covered by very high ratio, $>1$, gas. 
The very high ratios are predominantly from interarm regions and
appear to indicate the presence of wide-spread optically thin gas. 
Star formation tracers are better correlated with the
total neutral gas disk than with the molecular gas by itself
implying $\sdsfr \propto \sdg$. 
Using the 100 \fir\ and 21 cm continuum from NGC~6946
as star formation tracers, we arrive at a gas consumption timescale
of 2.8 Gyr, which is relatively uniform across the disk.
The high star formation rate at the nucleus appears to be due to a
large accumulation of molecular gas rather than a large
increase in the star formation efficiency.
The mid-plane gas pressure in the outer (R $>$ 10 kpc) HI arms 
of NGC~6946 is close to the value at the radial limit (10 kpc) 
of our observed CO disk. 
If the mid-plane gas pressure is a factor for the formation
of molecular clouds, 
these outer HI gas arms should contain molecular gas
which we do not see because they are beyond our detection limit.

\end{abstract}


\section{Introduction}

We have observed NGC~6946 as part of a program of deep mapping 
of extended, cold CO in nearby spiral galaxies \citep{CTHLMH01,CTBHM02}.
Large, fully-sampled, and deep images of CO in nearby galaxies are 
relatively uncommon. Interferometric images either do not contain zero
spacing data \citep[e.g.,][]{SOIS99} or if they do, the images typically 
do not go deep enough to detect cold, molecular disk gas. 
The detection of cold, extended molecular gas at large galactocentric 
radii in large spirals is a goal of this mapping 
project with the NRAO 12 Meter Telescope.

A gas-rich Sc galaxy at a distance of 6 Mpc 
\citep{ESK96, SKT97, KSH00}, NGC~6946 
is known for its bright and asymmetric optical spiral arms \citep{A66}. 
The pronounced asymmetry may be caused by interactions 
with neighboring dwarf galaxies, UGC 11583 and L149 \citep{SKT97,PW00}.
NGC~6946 has a 9\arcmin\ optical diameter on the sky and an
atomic hydrogen (HI) gas disk extending to 25\arcmin\ \citep{RSR73}.
Like many spirals, NGC~6946 has a central $\sim 2\arcmin$ diameter ``hole" in 
its HI disk.
Radio continuum, far infrared (FIR), optical line, and X-ray 
observations indicate vigorous star formation in the disk and an 
interstellar medium (ISM) stirred by supernovae and stellar winds 
\citep{E91, BV92, KS93, S94, LDG97}. 
The high level of star formation in the disk of NGC~6946 has been 
attributed both to its strong spiral density wave \citep{TY90}, and 
to stochastic, self-propagating, star formation \citep{DGSS84}.

Until recently, the CO morphology beyond the inner 3\arcmin\ radius 
($\sim$5 kpc) was not well known.
Like many spiral galaxies, NGC~6946 has a bright nuclear CO peak 
which falls off nearly exponentially with galactocentric radius 
\citep{ML78, YS82}. 
A 3\arcmin\ diameter map of \citet{TY89} revealed substantial CO emission 
in the inner disk with several emission peaks.  
\citet{CCVCB90} mapped two 2\arcmin\ circular fields in the
disk of NGC~6946 in a study of arm and interarm regions.
The inner 2\arcmin\ region has been mapped with higher resolution single dish
and aperture synthesis telescopes in transitions of 
CO and CO isotopomers \citep{BSSLS85, WCC88, SDINH88, Iea90, Wea93, RV95}.
A larger interferometer+single dish map of NGC~6946 has been made by the
BIMA SONG team \citep{Rea01}, although with lower sensitivity to cold
extended emission than the maps we present here. 
\citet{Wea02} presented 21\arcsec\ resolution fully-sampled single 
dish maps of CO(1-0).
We compare our results to theirs.  

The ``On The Fly"
(OTF) observing mode at the NRAO\footnote{\samepage
The National Radio Astronomy Observatory is a facility of the National
Science Foundation operated under cooperative agreement by Associated
Universities, Inc.} was ideally suited to the imaging of extended gas 
in galaxies.  Here, we present 16\arcmin\ by 10\arcmin\ maps of 
CO(1-0) and CO(2-1) 
covering the optical disk of NGC~6946 made with the 12 Meter Telescope.  
The maps are deep enough to detect cold, extended interarm CO to
levels of $\Ico \sim 1$ \iu, or $\Nmh \sim 2 \times 10^{20}$ \cdu.
We use these primary tracers of the molecular gas phase and combine 
them with archival and published data in order to study 
the molecular gas disk, the total neutral gas surface density, 
and their relationship to star formation in the disk of NGC~6946.

\section{The Observations and Data Reduction}
\subsection{12 Meter CO Observations}

The observations of CO(1-0) at 115 GHz and CO(2-1) at 230 GHz were
made at the NRAO 12 Meter Telescope at Kitt Peak, on two separate 
observing runs in December 1996 and March 1997.  An equivalent of 
$\sim$14 hours on-source observing was accumulated for each CO line.
Calibration was by the chopper wheel method \citep{UH76}. 
We convert the recorded \Trs\ values \citep{KU81} to main beam 
temperature, \TmbTrseqn, \nmb\ = $0.88\pm0.04$ at 115 GHz and 
\nmb\ = $0.56\pm0.06$ at 230 GHz \citep{M96a,M97}
\footnote{\samepage Available at http://www.tuc.nrao.edu/12meter/obsinfo.html.}.
We report $Tco = \Tmb$ throughout this paper.  
The filter bank spectrometer was configured for 2 MHz channel widths
with 256 total channels, producing spectral channels of a 5.2 \vu\ width
at 115 GHz and 2.6 \vu\ at 230 GHz.

The ``On The Fly" (OTF) observing mode was used \citep{M96b}
\footnote{\samepage Available at http://www.tuc.nrao.edu/12meter/obsinfo.html.}.
Scanning rates between 30 and 40 \arcsec/sec and scan row spacings of
18\arcsec\ at 115 GHz and 8\arcsec\ at 230 GHz were selected to ensure
sampling better than Nyquist over the mapped 
16\arcmin~$\times$~10\arcmin\ region.
The beamsize (FWHM) at 115 GHz is 55\arcsec\ and 27\arcsec\ at 230 GHz.
Chopper wheel calibration and sky offs were measured every 2 scans.
The rms noise level was reduced by averaging the 16 individual OTF 
maps made at 115 GHz and 10 individual OTF maps at 230 GHz.

The OTF data was gridded using the NRAO AIPS package.
A linear baseline was removed from each spectrum.
At this stage of the data reduction the background in the channel
maps is uneven from one scan row to the next.
This spatial variation, due to sky brightness fluctuations between 
successive off-source sky measurements, appears as ``stripes" 
in the continuum level running in the scan direction.
The maps were made large enough to give us emission free, 
2.5\arcmin\ wide, regions at the east and west ends of the maps, 
verified by the examination of individual spectra from these regions.
These regions were used to remove a linear baseline from each row of 
the channel maps in the scan direction, effectively giving us
a uniform background in all the channel maps. 
The rms variation in the channel maps before and after 
``de-striping" differed by 10 mK at 115 GHz and 30 mK at 230 GHz;
in both cases a 20\% reduction in the channel map noise.  

The mean rms noise ($1 \sigma$) from fully reduced, line free channels is 
\Tmb\ = 0.050 K and \Tmb\ = 0.145 K for the CO(1-0) and CO(2-1) respectively.
The channels were averaged to a 10.4 \vu\  channel width;
the mean rms noise in the averaged channel maps is 0.035 K for
CO(1-0) and 0.085 K for CO(2-1).   
The error beam of the 12m Telescope, which has been measured at 
345 GHz, is extrapolated to be 8-9\arcmin\, FWHM \citep{M97}. 
The contributions from the error beam 
in the individual channel maps is estimated to be less than a few 
percent, which we neglect here \citep[see discussion in][]{CTBHM02}.  
In the 10.4 \vu\ channel width cubes the rescaled uncertainty in 
\Tmb\ due to sky brightness fluctuations is $\pm7$ mK for CO(1-0) and 
$\pm11$ mK for CO(2-1). 

We are interested in bringing out as much of the extended low level
CO emission as possible in our integrated intensity maps.
To do this we want to minimize the noise that would
normally be added to an \Ico\ map 
when the channel maps are added together.
To test the effects of the level of flux cut (level in the individual 
channels below which the data is deemed to be noise and not added 
to the integrated map) on the final integrated flux, we did 
the following test.  
We convolved the CO cubes to twice the beam size and used 
the regions with emission  $>3 \; \sigma$ to create mask cubes. 
The mask cubes allow us to define areas of emission using smoothed, 
hence more sensitive images, which is one way of discriminating between 
signal and noise. 
We then created \Ico\ maps from both the masked and unmasked cubes
using flux cuts from 1.0 $\sigma$ to 0 $\sigma$ in 0.1 $\sigma$ decrements. 
The total flux in the CO(1-0) integrated intensity map made from 
the masked cube with a 0.5 $\sigma$ flux cut agrees with flux from the
unmasked cube with no flux cut, to within 1\%. 
For the CO(2-1) map, a more conservative 0.8 $\sigma$ was used.

We created maps of the per pixel uncertainty for our \Ico\ maps
by combining the per channel sky brightness fluctuation and 
rms noise, counting the number of channels contributing to the 
integrated emission at each pixel, and then adding the uncertainty
in \nmb\ using error analysis techniques for the sums and products 
of independent variables.
We created maps of the fractional uncertainty in our integrated
intensity maps by dividing the aforementioned uncertainty maps
by the \Ico\ maps themselves.
The uncertainty is dominated at the nucleus by the uncertainty in \nmb\,
while at the outer edges of the maps the uncertainty is dominated by
the rms noise.
For the CO(1-0) integrated intensity map the combined uncertainty 
is on the order of 7\% at the nucleus, 12\% along 
the prominent spiral arm pattern, and 25\% for the outer disk emission.
For the CO(2-1) map the uncertainties are on the order of
11\% at the nucleus, 20\% along the spiral arm
pattern, and 30\% for the outer disk emission.

To obtain $\rm H_2$ column densities  from \Ico\ we
applied the  ``standard conversion factor", \Xco\ \citep{SS87, YS91}.
\Xco\ is estimated to be accurate to a factor of two in
the Milky Way disk \citep[][and references therein]{SRBY87}, although
it may overpredict \Nmh\ in the central arcminute of large spirals 
\citep{DHWM98,MT01}. We assume that these uncertainties
hold for NGC~6946 as well.
We will adopt the standard conversion factor, $$\Xcoeqn,$$ of
\citet{Sea88} throughout the remainder of the paper
\citep[a recent calibration by]
[finds a mean \Xcoval{1.6} for the entire Milky Way disk and 
$\Xcoval{2.7}$ for the outer disk, R $>$ R$_{\sun}$]{Hea97}.
Using this conversion factor, \Ico\ can be converted into \mh\
(no He or heavier elements) surface densities using
$$\sdmh (\sdu) = 2.4\: \times (\Ico /\; \iu)$$ including a geometric
correction for inclination of the galaxy (40\arcdeg, Table 1)
\footnote{\samepage The adopted inclination was derived from
a Brandt model fit to the HI data.}
to correct the surface densities to their face-on values.

\subsection{HI, 21 cm Continuum and FIR Maps}

We obtained archival VLA D array 21 cm observations of the HI
line in NGC~6946, which are described in \citet{TY86}.  The beam
size is 49\arcsec~$\times$~42\arcsec, with $pa = 73 \arcdeg$, 
and each channel has a velocity width of  10.3 \vu.
The integrated intensity map was constructed from signals greater 
than 1.7 \mjyb\ (1.2 $\sigma$). 
Absolute flux calibration error is on the order of the 
uncertainty in the flux of 3C~286, $< 5\%$. 
The single channel rms noise of 1.4 \mjyb\ corresponds to an HI column 
density of $7.7 \times 10^{18}$ \cdu. 

Integrated intensities were converted to face-on HI surface densities
(no He or heavier elements) using, 
$$\sdhi = 3.3\; \sdu\: (\su)^{-1}\; \cdot \Ihi .$$
From a comparison to the single dish observations of \citet{GRR68}
we estimate 40\% of the HI flux of NGC~6946 is missing from the VLA maps
due to the lack of short baselines (emission extended over 15\arcmin\ ). 
If the missing flux exists as a screen across the disk of NGC~6946 it 
constitutes a missing uniform HI surface density of $\sim 1.2$ \sdu.

The total observed HI emission in the VLA maps amounts to 
$\Mhi = 5\times10^{9}$ \Mo, not including missing under-sampled emission. 
\citet{TY86}, using the same VLA HI data, obtained
$3.9 \times 10^{9}$ \Mo, rescaled to D = 6 Mpc. 
Their mass is probably slightly lower than ours because they used a 
higher flux cutoff to produce their integrated intensity map.
\citet{BV92} obtained $6.7 \times 10^{9}$ \Mo\ (rescaled to 6 Mpc).
Boulanger's mass includes single dish observations merged with
the Westerbork interferometer data to cover the short spacings and
corrections for Milky Way absorption and emission.  The difference
between our mass and theirs, 33\%, is in line with our estimate
of the missing flux in the VLA map.

A Brandt rotation curve was fit to the HI radial velocities,
for radii $< 10 \arcmin$. 
Our resulting kinematic parameters, Table 1, are in good agreement 
with the rotation parameters found by \citet{CCBV90}. 
A total dynamical mass estimate based on the fit
(see Table 1) is: $M_{dyn} = 1.9(\pm0.6) \times 10^{11}$ \Mo.

We produced a 21 cm continuum map by averaging  4 ``line free" 
channels in the NGC~6946 HI channel cube.   
The $1 \sigma$ noise level is 1.2 \mjyb. 

High resolution (HiRes) 60 and 100 \fir\ IRAS maps were obtained from IPAC
\footnote{\samepage Descriptions of the IRAS HiRes data reduction and 
HiRes data products are available at 
http://www.ipac.caltech.edu/ipac/iras/toc.html.}.
The 100 \fir\ HiRes map was produced by 200 iterations of a maximum 
entropy method yielding a 69\arcsec~$\times$~65\arcsec, 
$pa = 20 \arcdeg$ beam and a $1 \sigma$  noise level of 2.0 \Mjys.
The 60 \fir\ HiRes map was produced by 50 iterations of a maximum entropy 
method yielding a 45\arcsec~$\times$~41\arcsec, $pa = 21 \arcdeg$ beam 
and a $1 \sigma$ noise level of 0.6 \Mjys.

\section{Images of CO in NGC~6946: The Molecular Gas}

A relatively heavy molecule, CO is easily excited at the 
temperatures of molecular clouds, with its first rotational 
level at an energy of E/k = 5.5 K. 
Since it has a small dipole moment and is generally optically thick, the 
excitation of the CO rotational levels is driven by 
collisions at \mh\ densities above $n_{crit} \sim 300$ \du\ for 
the J=1-0 transition. The J = 2-1 transition, with an upper level energy of
E/k = 16.6 K,  has a higher critical density, although
its higher optical depth somewhat compensates for that. 
For gas kinetic temperatures of $\sim$7-10~K, we expect CO(2-1) to be
thermalized at critical densities above $n_{crit}\sim 10^3$ \du,
slightly  higher than CO(1-0) values.
One might therefore expect  CO(2-1) emission to be more
tightly confined to the spiral arms than CO(1-0), since the arms probably 
have denser gas, and are also closer to HII regions that might 
warm the clouds \citep[e.g.,][]{SY87}.
However, as we shall see, the CO(2-1) and CO(1-0) properties in 
NGC~6946 are not precisely what one might expect based on Galactic clouds.

We do not present CO kinematic data (radial velocity maps, 
velocity dispersion maps, rotation curves). Our kinematic results
do not differ significantly from those of \citet{Wea02}, and we
refer the reader to their presentation of the kinematics of
NGC~6946 derived from IRAM single dish data. 

\subsection{CO(1-0) Maps}

The 24 CO(1-0) channels containing line emission are displayed in 
Figure \ref{co10.chnl}.  
The channels have been smoothed to a $10.4\; \vu$ channel width.
The channel maps show the butterfly pattern characteristic of emission
from an inclined rotating disk.
High velocity nuclear gas shows up as emission at the location of the 
nucleus spread across nearly the entire bandwidth of the line emission.

The disk of CO emission in NGC~6946 extends 8\arcmin\ (14 kpc) 
north-south by 10\arcmin\ (18 kpc) east-west as shown in the  
integrated intensity, \Iozeqn\ (Figure \ref{co10.m0tb}a), 
and the peak main beam brightness 
temperature, \Toz, (Figure \ref{co10.m0tb}b) maps. 
\citet{Wea02} also found CO emission extended out to 10\arcmin\ 
in 22\arcsec\ resolution single dish IRAM observations. 
There are no significant differences between our and Walsh's
map that cannot be attributed to the difference in resolution.
The CO(1-0) emission covers the entire optical disk of a
Digital Sky Survey (DSS) image of NGC~6946 (Figure \ref{co10.m0tb}c).

The total CO(1-0) flux density measured over a velocity width of 
$250\; \vu$ is $1.2(\pm0.3) \times 10^{4}$ \sut, in agreement with the 
value published in the FCRAO survey of $1.2(\pm0.4) \times 10^{4}$ \sut\
\citep{Yea95}, but higher than the $0.79(\pm0.01) \times 10^{4}$ \sut\
detected in a 22\arcsec\ resolution single dish IRAM map \citep{Wea02}.
Converting our total observed \Ioz\ into a total molecular mass, we 
obtain $\Mmh = 3.3 \times 10^{9}$ \Mo\ for NGC~6946; 
the mass will be higher if there is outer disk CO that we do not detect.  
This is consistent with the mass obtained by \citet{TY89}, 
$3.0 \times 10^{9}$ \Mo\ (rescaled to 6 Mpc and our $X_{CO}$).  

The CO integrated intensity (\Ioz) and temperature maps (\Toz) both 
show the same features: a bright
nucleus, a broad asymmetrical spiral arm pattern and the overall
asymmetrical distribution of the disk emission.
The nucleus has a CO emission peak at 
\rah{20}\ram{34}\ras{52.6}\ \decd{60}\decm{9}\decs{16} (J2000), 
with $\Ioz^{nuc}= 72$ \iu,  $\Nmh^{nuc} = 1.4 \times 10^{22}$ \cdu,
or in terms of surface density, $\sdmh^{nuc} = 170$ \sdu\ (face-on value), 
averaged over our 1.6 kpc diameter beam.  
The total molecular mass within the central beam is $4.7\times 10^{8}$ \Mo.  
\citet{Iea90} obtained $4.5\times 10^{8}$ \Mo\ (adjusted to 6 Mpc 
and our conversion factor) in a 65\arcsec\ diameter region from 
interferometric observations that recovered 70\% of the total flux. 
Large scale CO spiral structure is traced in a broad ``S" pattern, 
outlined by the 7.5 \iu\ contour in the \Ioz\ map and more clearly 
seen in the \Toz\ map, outlined by the 0.23 K contours. 
Here, molecular gas surface density is $> 20$ \sdu. 
There is also extended, weaker CO emission from the disk outside of the
7.5 \iu\ contour in the \Ioz\ map.
Large regions of molecular gas exist to the northwest and southeast 
of the nucleus, as well as patchy emission out to the edges of the maps.  
Not all of this gas is associated with the weaker optical arm pattern.
The face-on gas surface density for this extended region has a mean
\Ico\ of  $\Ioz \sim 3$ \iu\ corresponding to 
$\rm N_{H_2} = 5 \times 10^{20}$ \cdu\ or $\sdmh \sim 10$ \sdu.
This extended portion of the gas disk contributes 1/4 of the total CO 
emission we observe. The total mass of this extended gas disk 
is $\Mmh \sim 10^{9}$ \Mo.  

\subsection{CO(2-1) Maps}

The 24 channels containing CO(2-1) emission are displayed in 
Figure \ref{co21.chnl}.  The rms of these  10.4 \vu\ channels is three times
as high as for CO(1-0), so the maps are less sensitive to faint emission.
The total CO(2-1) flux density measured over a velocity width of 
$250\; \vu$ is $3.5(\pm1.0) \times 10^{4}$ \sut.
The butterfly patterns for the CO(2-1) and CO(1-0) are generally the same, 
with the differing resolutions, 27\arcsec\ and 55\arcsec\ respectively, 
accounting for most of the differences. 

As with CO(1-0) the bright nucleus and asymmetric distribution 
of arm and disk emission are also apparent in the integrated
intensity, \Ito, and peak main beam brightness temperature, \Tto,
maps (Figures \ref{co21.m0tb}a and \ref{co21.m0tb}b).
The CO(2-1) emission in NGC~6946 is almost as extensive as that 
of the CO(1-0): the north-south and east-west dimensions are only 
1\arcmin\ shorter. 
The observed CO(2-1) emission covers the brighter portions of
the DSS optical image of NGC~6946, Figure \ref{co21.m0tb}c.

The nuclear peak of the  CO(2-1) integrated intensity map is
at \rah{20}\ram{34}\ras{53.1} \decd{60}\decm{9}\decs{15} (J2000) 
with $\Ito^{nuc} = 125$ \iu. 
Using a Galactic disk $X_{CO}$, this corresponds to a peak column 
density of $\Nmh^{nuc} = 2.2\times10^{22}$ \cdu\ 
or 450 \sdu\ averaged over the 790 pc beam. 
If other nearby galaxies are any indication,
this may be an overestimate of the true molecular gas mass in the 
nucleus, by factors of $\sim 2-3$($\S$3.1.) 
The largest \Tto\ value, 0.85 K, occurs at the same location as the \Ito\
peak and both peaks are consistent with the location of the \Ioz\ peak. 

At the resolution presented of maps, inside a 2\arcmin\ (3.5 kpc) 
radius both the CO(2-1) and CO(1-0) emission trace a broad 2 arm pattern.
At a 2.5\arcmin\ radius (4.4 kpc) the CO(2-1) still traces a
broad 2 arm pattern while the CO(1-0) emission from a larger beam
has clearly bifurcated into a 4 arm spiral (Figure \ref{coaz}a).
At a 3.5\arcmin\ radius (6.1 kpc) 5 arms are discernible in
azimuthal emission plots of both CO emission lines (Figure \ref{coaz}b).

Along the optical arm pattern the CO(2-1) emission is uneven, 
(Figure \ref{co21.m0tb}c) with typical variations of 
15 and 5 \iu\ separated by 40\arcsec\ or less.
This same clumpy pattern of CO arm emission can be seen in the 
22\arcsec\ CO(1-0) and CO(3-2) maps presented by \citet{Wea02}.
The emission on each arm at the same galacto-centric radius is also
uneven as can be seen in Figures \ref{coaz}a and b.

At 27\arcsec\ resolution it becomes possible to distinguish between
arm and interarm emission. 
The \Ito\ contrast between the brighter, on arm, clumps and obvious 
off-arm regions is $\sim 3$. 
Using the DSS optical image to mask the outer (R$>$ 2.5\arcmin) 
arm/interarm regions,
we obtain a mean arm/interarm contrast of 2.4. 
This is higher than both the 1.2 and 1.8 values found from 
CO(1-0) and CO(3-2) IRAM observations of \citet{Wea02} who 
interpret the increase in contrast at the higher CO transitions 
verses CO(1-0) as evidence that the on-arm molecular gas is warmer.

The presence of widespread interarm gas becomes plainly apparent
in a comparison to an optical image, Figure \ref{co21.m0tb}c.  
The fact that interarm CO(2-1) is present and at emission levels
comparable to CO(1-0) suggests that the interarm gas is also 
not very cold and that it must be warmer than $\sim4-7$ K.  
This widespread distribution of interarm gas is not seen in the
\citet{SHHHO97} CO(2-1) survey of the Milky Way. 

\subsection{Excitation of CO in NGC~6946}

A CO ratio map, $\Rco = \Ito/\Ioz$, is presented in Figure \ref{Rco}a.
The CO(2-1) channel cube was convolved to the beam size of the CO(1-0) 
data and the $>3\sigma$ emission from the beam matched cubes was used to
produce the \Rco\ map.
The spiral pattern is  only weakly reflected in the \Rco\ map. 
Outside of the inner few arcminute radius the pattern breaks up 
into a patchwork of high and low \Rco\ regions.  

The ratio uncertainty for the regions of strong emission is dominated 
by the error in the conversion to main beam temperature, while
the uncertainty for the weaker emission regions is dominated
by the rms noise in the contributing channels of emission.
Adding in quadrature the fractional uncertainties from the \Ico\ maps 
we obtain an error estimate on the order of 15\% at the nucleus, 
25\% along the spiral arm pattern, and 40\% for the outer disk ratios.
We were concerned that the ratio might be biased towards higher 
values due to the uncertainty in the denominator. 
We performed a standard statistical analysis to correct a ratio
for bias this bias.
The bias corrected \Rco\ map was not significantly different from the 
map presented in Figure \ref{Rco}a and well within our reported error.
Despite the large uncertainty estimates, our ratios are consistent
with those derived from other independent observations 
\citep{WCC88,Iea90,CCVCB90,Wea93}.

In principle, \Rco, the ratio of CO(2-1) to CO(1-0) integrated intensities,
gives information on the excitation of the CO emission and depending 
on circumstances, the kinetic temperature, \Tk.
The observed main beam temperature at frequency, $\nu$, is related
to a Planck function, \Bu, with excitation temperature, \Tex, by the
areal beam filling factor, \fa, a conversion to the Rayleigh-Jeans
temperature scale, and radiative transfer through a optical
depth, \tu : \Tmbeqn
including a correction for the cosmic microwave background contribution
to the beam, \Bucmb, at temperature 2.73 K.
Standard assumptions are that \fa\ is the same for both emission lines,
that both lines are optically thick ($\tu \gg 1$) and that the
emission can be characterized by a single \Tex\ 
\citep{DSS86, MB88, S96}.  
With these assumptions we can naively derive \Tex\ values for 
the emission: \Rco\ = 0.9 corresponds to \Tex\ = 20 K, 
while \Rco\ = 0.5 corresponds to \Tex\ = 3.5 K.
Higher ratio values are found in the nuclear regions of
starburst galaxies where elevated star formation heats the molecular
gas; lower values are typical of cold or subthermally excited
(\Tex $<$ \Tk) disk molecular clouds.
Values of $\Rco > 1$ indicate optically thin gas in this picture.

Within 1\arcmin\ of the nucleus, $\Rco \sim 0.8$.  This is
consistent with other studies \citep{WCC88,Iea90,Wea93} 
that indicate the presence of relatively dense 
molecular gas as temperatures of about $\Tex \sim 10-15$ K.  
This is comparable to the Milky Way mean \Rco\ of 0.74 
\citep{OHHHS96,OHHHS98}.

Along the optical arms the mean \Rco\ is 0.83, ranging 
from 0.44 to 1.4.  
To within our uncertainties, this is the same as the 0.73 mean 
found for Milky Way molecular arms where a similar range of \Rco\ 
values is found \citep{SHHHO95,SHHHO97}.  
From a naive interpretation of the ratios, a mean \Tex\ of 12 K is
implied for the arm CO in NGC~6946.

\citet{CCVCB90} concluded that the interarm \Rco\ does not 
differ significantly from that of the arms based on observations 
of two circular regions in the arms, one on the east side
of NGC~6946, the other on the west side. 
The two circular regions they observed have a mean \Rco\ of 0.8
in our ratio map, also not significantly different from that of 
the arm regions.
However when larger regions of the CO disk are sampled, a very
different picture emerges. 
There are regions (5\% of the ratio map) where $\Rco < 0.6$ 
indicating cold ($\Tex <$ 5 K) or sub-thermally excited CO. 
There are also much larger regions where $\Rco > 1$. 

Very high ratio, $\Rco > 1$, gas covers 50\% of our ratio map; 
and these are predominantly interarm regions, Figure \ref{Rco}b. 
We have eliminated the possibility that these very high ratios are 
due to contributions from the telescope error beam at 230 GHz:
the error beam at the 12 Meter Telescope is a measured quantity
and modeling of the error beam contribution in the individual 
channel maps shows that it contributes at most a  few percent of the 
emission in the \Ito\ map.

The high \Rco\ ratios appear to indicate the presence of widespread,
warm, optically thin CO-emitting gas in the interarm regions.
While optically thin CO emission is not typical of Galactic molecular
clouds, our large beam and unusual perspective from outside the
disk of NGC~6946 may allow us to trace a component that is difficult 
to detect in the Galaxy. 
We note that the photon-dominated regions (PDR) around cool stars can
produce bright CO emission with high temperatures \citep{Sea94}
since their softer radiation fields allow the CO to be
photoelectrically-warmed at low A$_v$.
Also, as pointed out by \citet{WRHB90} it does not take much 
optically thin gas to swing the emissivity; the CO(2-1) emission
that we see could easily come from a small amount of optically
thin, but ``CO-loud" gas that is not the dominant cloud population.
Other less likely possibilities are discussed in \citet{CTBHM02}.

We have also seen a have a high overall $\Rco \sim 1$ in M83 
\citep{CTBHM02}; confirming earlier reports of unusually high 
\Rco\ in these two galaxies \citep{Cea90,WRHB90}.  
Large scale \Rco\ maps of the Milky Way
\citep{SHHHO95,OHHHS96,SHHHO97,OHHHS98} show high ratios in
immediate proximity to pockets of star formation. 
The widespread regions of high \Rco\ seen in M83 and NGC~6946 are not 
seen in surveys of CO in our Galaxy. 

\section{CO, HI and the Neutral Gas Surface Density}

Although CO emission covers only 15\% of the HI disk surface area,
NGC~6946 is nevertheless a molecular gas-rich galaxy; 
1/3 of the hydrogen gas disk is in molecular form. 
This is consistent with the 36\% value from \citet{Wea02} once the flux 
missed in the interferometer HI maps is taken into account. 

In Figure \ref{hico}a, we present our CO(1-0) emission in contours 
superimposed on a grey scale image of the HI emission.  
This image shows that CO emission fills the inner 6\arcmin\ of the HI disk. 
The HI disk has a central depression, which rises to peaks in a clumpy 
ring of emission. 
The HI ring is coincident with the outer boundary of CO(1-0).  
While spiral structure is readily apparent in the outer HI disk \citep{TY86}, 
it is insignificant in the inner disk, R$< 5 \arcmin$\ (9 kpc), 
where the HI column densities decline and $H_2$ takes over.  
At 1\arcmin\ resolution there are no obvious systematic displacements of the 
HI relative to the CO(1-0) spiral structure other than the strongest HI 
tends to be near the outermost ends of the CO(1-0) arms.  
In fact, if anything, HI and CO peaks tend to
coincide, which is more apparent in the higher resolution  CO(2-1) maps.

The CO(2-1) image in contours is shown in Figure \ref{hico}b, again 
with a grey scale image of HI. 
Outside the inner 2\arcmin\ (3.5 kpc) region of the HI depression, 
the peaks of CO(2-1) emission tend to coincide with HI peaks 
along the ring.  
This could be evidence for photodissociated \mh\ creating a 
higher HI column density at these locations. 
\citet{TA89,TA93} and \citet{WS91} suggest the same explanation for HI 
in the spiral arms of M51, M83 and M33.  
 
From these images, we have constructed plots of the azimuthally-averaged
gas surface densities, \sdmh, \sdhi\ and \sdg, 
 as a function of galactocentric radius in Figure \ref{sdg_r} 
(corrected for inclination). 
This is similar to the plot of \citet{YS82}, which was based on a radial 
cross map, and those of \citet{TY86} and \citet{Wea02}.
\mh\ dominates the neutral gas distribution in within the central
3\arcmin\ (5 kpc) of NGC~6946, 
rapidly fading into a predominantly HI disk beyond. 
A least squares fit of an exponential to  \Ico\
over the entire range of radii has a scale length of $1.0\arcmin \pm0.1$ 
(1.8 kpc). 
\citet{TY86} and \citet{Wea02} obtained scale lengths of 1.5\arcmin\
and 1.4\arcmin\ for the CO disk from FCRAO and IRAM single disk observations; 
\citet{Rea01} obtained 1.3\arcmin\ from the BIMA interferometer survey data
combined with NRAO 12m observations. 
All these latter scale lengths exclude the nuclear contribution to 
the fit and are sensitive to the range of radii over which they computed.
We obtain $1.2\arcmin$ and  $1.3\arcmin$ fitting to the range of radii used 
by \citet{TY86} and \citet{Wea02}, 40\arcsec\ - 280\arcsec\ 
and 60\arcsec\ - 260\arcsec\ respectively. 
We conclude that the CO falls off with an exponential scale length
of $1.2\arcmin \pm 0.2\arcmin$ on arcminute scales, excluding the 
inner 1\arcmin\ nuclear region and about 1\arcmin\ if the nucleus
is included.

A total neutral gas surface density map, 
\sdgeqn $\:$  is presented in Figure \ref{sdg}a.  This map has been
corrected for inclination but not for missing \sdhi\ due 
to under-sampling in the VLA image ($\sim$1 \sdu),
A false color image is presented in Figure \ref{sdg}b, with HI in red and
CO in green, so that the 
molecular contribution to the total gas disk can be distinguished.
Values for \sdg\ range from 240 \sdu\ at the nucleus,
to 20-45 \sdu\ along the spiral arms within the optical part of the galaxy,
to 5-10 \sdu\ for the outer arms.
At first glance, \mh\ and HI blend into a single
global gas structure extending from the nucleus to the outer gas disk. 
A closer look reveals some subtleties. 
In the southern part of NGC~6946, the neutral density gradient is smooth 
and gradual. In the north, the gradient is much steeper, with the gas 
density falling to a level of $20$ \sdu\ in 90\arcsec\ 
(2.6 kpc) to the north, into an interarm void, as compared to a 180\arcsec\
fall to the same mass density in the south (this can also be seen 
in Figures \ref{hico}). 
This steeper falloff on the northern side of the central gas peak may 
be related to the more well-defined spiral arm to the north, and perhaps 
also to the prominent northeastern optical arm \citep{A66}.

The global properties of three late galaxies for which we have
a similar data set, along with the Milky Way for reference are 
listed in Table 2.  
IC~342, M83 and NGC~6946 all have gas disks that extend well beyond 
their optical diameters and all three show some evidence for a bar. 
The $\sdmh^{disk}$ values for IC~342, NGC~6946 and M83 are all 
significantly higher than that of the Milky Way;
a factor of four higher for IC~342 and NGC~6946, 
a factor of seven higher for M83.  
The NGC~6946 $\sdmh^{disk}$ value is similar to that of IC~342, 
while its $\sdmh^{nuc}$ intermediate between IC~342 and M83.
All four galaxies have similar relative proportions of \mh\ and HI,
global \Mmh/\Mhi\ values between 0.3 and 0.5 are also seen
in the nearby galaxy, Maffei 2 \citep{MW04}.
M83 stands out as a particularly gas-rich object relative to its 
dynamical mass, with a factor of two higher M$_{gas}$/M$_{dyn}$
than IC~342 and NGC~6946, and four times that of the Milky Way.  
The \mh\ and optical disk diameters of the three external
galaxies are the same, D$_{H_{2}}$/D$_{25} \sim 1$.
The dynamical masses of these three galaxies are an order of 
magnitude smaller than that of the Milky Way, although their total 
gas mass is similar.  
It is difficult to place too much emphasis on subtle variations
that rely on distance estimates, since the uncertain distances to 
these nearby galaxies can easily affect their masses, both gaseous 
($\propto R^2$) and dynamical ($\propto R$).

\section{Star Formation and the Neutral Gas Disk of NGC~6946}

\subsection{Star Formation Tracers, CO, and the Neutral Gas Disk}

Because stars form from molecular gas, we expect to see correlations 
between CO and tracers of star formation. These tracers include: 
\ha\ emission from ionized gas associated with massive young stars,
FIR emission from dust heated by young stars,
and radio continuum emission from relativistic electrons created by 
supernovae.  
To calculate linear correlation coefficients all the maps were
convolved to the largest beam size in our collection of maps, 
a 70\arcsec\ circular beam size, { \it then clipped to use $>3\sigma$
emission}, and sampled on a 35\arcsec\ grid. 

\ha\ emission \citep{FWGH98} and \Ito\ are compared in Figure
\ref{cosf}a.  
The nuclear bar traced by \ha\ within the central 2\arcmin\
is aligned with the elongation of the central \Ito\ contours. 
The \Ito\ arms are traced at higher resolution by the \ha\ emission.
In the outer \Ito\ disk, the brighter patches of CO(2-1) emission
are aligned with the brightest patches of \ha\ which is most
clearly seen along the northeastern spiral arms.  This supports
the conjecture that the ring of HI emission seen at these 
radii is the result of photodissociated \mh. 
It is apparent from this map that ratio of \ha\ emission
to \Ito\ is not the same in the nucleus and outer arms;
due to substantial extinction particularly in the nucleus, 
\ha\ is probably not the optimal 
tracer of star formation for this gas-rich galaxy.
When \ha\ is compared to \Ito\ or our primary molecular mass tracer, \Ioz,
we obtain a linear correlation coefficient of $\sim 0.65$.
The \ha\ correlations do not improve when a 1\arcmin\ diameter region centered
on the nucleus is excluded from the correlation.

Figure \ref{cosf}b shows that the 60 \fir\ FIR emission generally 
reproduces the morphology seen in \Ito\ (grey scale), 
despite the noticeable artifacts (the ``boxy" structure of the
central contours) produced by the HiRes maximum entropy algorithm. 
An ISOPHOT 60 \fir\ map by \citet{Tea96} confirms the overall 
morphological structure seen in the HiRes map.
The 60 \fir\ nuclear peak is at \rah{20}\ram{34}\ras{52.4} 
\decd{60}\decm{9}\decs{15} (J2000) which is, within the uncertainties,
coincident with the CO peak. 
If CO(2-1) along with 60 \fir\ FIR trace a warmer component both 
heated by star formation; we expect these emission pairs to 
have similar feature in the disk. 
The 60 \fir\ emission traces the overall CO(2-1) arm structure
better than the 100 \fir\ emission shown in Figure \ref{cosf}c.
The 4 prominent peaks in the outer 60 \fir\ emission disk
are aligned with peaks in the outer \Ito\ arm structure.

The 100 \fir\ FIR and 21 cm continuum maps have resolution comparable 
to our CO map (Figures \ref{cosf}c and d).
The 100 \fir\ emission roughly mimics the CO(1-0) emission disk,
but the CO arm structure east of the nucleus is not well traced
by the 100 \fir. 
The 100 \fir\ nuclear peak is at \rah{20}\ram{34}\ras{53.0}
\decd{60}\decm{9}\decs{23} (J2000); northeast of the CO emission peaks.
The radio continuum does a better job reproducing the CO morphology, 
with the exception of the 21 cm peak 4\arcmin\ northwest of the nucleus. 
This is consistent with the suggestion that cosmic rays may be 
more important for the heating of molecular gas on large scales 
such as observed here
than the radiation from in situ massive star formation \citep{AAL91,SAH93}.
The continuum peak to the north and west of the nucleus at 
\rah{20}\ram{34}\ras{24.9} \decd{60}\decm{10}\decs{38}  is probably a 
background radio galaxy \citep[][based on its spectral index, 
$\alpha = -0.84$ and flux.]{KAR77}
The 21 cm continuum nuclear peak at \rah{20}\ram{34}\ras{52.4} 
\decd{60}\decm{9}\decs{15} is consistent with the CO(1-0) peak.
Much of the outer CO arm morphology and prominent emission
peaks are traced by non-thermal radio continuum.

\citet{Wea02} also found statistically significant correlations 
between CO(1-0) emission and the total radio-continuum at 6 cm, 
and between CO(1-0) and 20 cm non-thermal continuum in NGC~6946.
When fluxes are extrapolated to the expected 20 cm values \citep[using 
$S_{\nu} \propto \nu^{-0.8}$]{C92}, we obtain a mean 
\Ioz/T$_{20}$ of $1.4\pm0.5$ \vu\ 
from our 70\arcsec\ beam convolved maps for the disk of NGC~6946.
This is higher that the \Ioz/T$_{20}$ =1.0 value determined by 
\citet{AAL91} from sparely sampled \Ioz\ data, but in line with the 1.3
mean value for all the galaxies in their sample. 

One might naively think that the FIR has a more direct causal link 
to newly formed stars, while radio continuum emission, which arises 
from relativistic electrons produced in supernova remnants,
would be more spatially removed. 
In reality the far-infrared cirrus and non-thermal disk emission 
on arcminute scales represent a less temporally localized reflection 
of star formation;  the close correspondence of radio continuum and CO
supports a model of cosmic ray confinement on 1 kpc scale lengths. 
and highlights the importance of gas surface density in containing
the magnetic field required to extract synchrotron emission from
relativistic electrons \citep{BHC89,BH90,HB93}.

Table 3 lists least squares fits to the radial distribution of ratios
of \Ioz\ or \Ito\ to our star formation tracers, \gasSF. 
With the exception of \ha, 
all the star formation tracers are highly correlated with CO emission.
From the 70\arcsec\ beam maps we obtain linear correlation coefficients
on the order of $~\sim0.95$ or better in comparisons between either of
our CO maps and any of the star formation tracers except for \ha.
The radial gradient is reduced by a factor of two when \Ioz\ is
used instead of \Ito\ indicating \Ioz\ remains the best tracer
of molecular gas associated with star formation on kpc size scales.

While CO is highly correlated with star formation tracers, the 
correlation of total gas, HI$+$CO, is even better (see Table 3).
The radial gradient is reduced by a factor of ten when \sdg\ is used.
If we assume that the 21 cm and FIR emission trace star formation, 
this result implies that the star formation
rate in NGC~6946 is more closely tied to \sdg\ than \sdmh.
Restating this result: a simple Schmidt law applies, SFE $\propto \sdg^n$ 
\citep{S59,K98b}, at least on kpc size-scales. 
{\it Moreover, it appears that in the outer parts of NGC~6946, star
formation is likely to take place in regions of predominantly HI gas}.

We obtained similar results for IC 342 and M83 \citep{CTHLMH01, CTBHM02}.
While the correlation coefficient for \sdg/\ftocm\ in M83 is also 
high, 0.95, for IC~342 the correlation coefficient is only 0.73.
The lower map correlation for IC~342 is related to the peculiar
1 armed appearance of the 21 cm continuum map that contrasts starkly
with its grand design appearance in other tracers.  
From least squares fits to the \sdg/\ftocm\ ratio we obtain slopes 
of $-0.012\pm0.004$ (dex) for IC~342 and $-0.13\pm0.02$ (dex) for M83. 
For M83, the slope is not as flat as in IC~342 and NGC~6946.  
This gradient in M83 may indicate an age difference with larger 
\sdg/\ftocm\ values resulting from younger star forming regions 
in its disk \citep{SAH93}.
The \sdg/\ftocm\ comparison for NGC~6946 suffers none of these 
peculiarities.

A complication in the interpretation of molecular gas mass could 
arise because of a radial gradient in metallicity and because
the Galactic CO conversion factor over-predicts the \mh\ mass 
in the nucleus of NGC~6946 \citep{MT04}.
Based on the metallicity gradient for NGC~6946
measured by \citet{BR92} and the metallicity adjustment to
the CO conversion factor used by \citet{W95} for giant molecular
clouds in M33, \Xco\ could be a factor of two higher at the edge of
the NGC~6946 CO disk.  This would increase the molecular surface
density derived from CO at the disk edge by a factor of two, and 
potentially reduce the radial gradient of molecular mass implied
by \Ioz\ with radius. 
However, this variation in \Xco\ does not appear to be universal 
in low metallicity systems \citep{MTB02,LBSB05}, so we do not adopt the
correction here.

\subsection{The Star Formation Efficiency of the Disk of NGC~6946}

Studies of the links between star formation and gas, or star formation 
efficiency, have recently focused on correlations as a function of 
position in the galaxy, as opposed to global measures.
\citet{RY99} examined the relation between \sdmh\
and the star formation rate, SFR, as traced by \ha; 
\citet{MCMG02} compare a radio continuum based SFR and \sdmh; 
\citet{WB02} compare a SFR from \ha\ emission to \sdg.
\citet{YS82} pointed out that the falloff of \sdmh\ in NGC~6946
with radius was very similar to the falloff in B luminosity, implying 
a constant star formation rate per unit H$_{2}$.
The most widely used formulation is the Schmidt law, 
$\sdsfr = A\, \sdg^{N}$ \citep{S59}; and the most widely accepted 
calibration was done by \citet{K98a} who finds 
$A = 2.5\pm0.7 \times 10^{-4}$ and $N = 1.4\pm0.15$ for \sdg\ in 
\sdu\ and \sdsfr\ in \sdsfrkpcu.  
How well does the Schmidt law describe star formation in NGC~6946?

We can estimate star formation rates from our FIR and 21 cm continuum data.  
Unlike \ha\ emission, these tracers are relatively unaffected by absorption.
Although cirrus heated by an older population of stars may contaminate 
FIR in some galaxies, this is  probably not a big effect in Scd type spirals
\citep{ST92}. 
The 21 cm continuum appears to be a good star formation tracer on global 
scales, although it must break down at smaller scales.  
The maps for this analysis were convolved to the the largest circular 
beam in the set, 70\arcsec.  

We built a FIR map from the 60 \fir\ and 100 \fir\ IRAS maps using: 
$$ S_{fir} \: (\Wms) = 12.6 \times 10^{-14} \; 
   (2.58 \: S_{60 \mu m} \: (\jys) + S_{100 \mu m} \: (\jys)) $$
\citep{CGQ85}.  We adopt the FIR star formation rate:
$$ \sfrfir \: (\sfru) = 4.5 \times 10^{-37} \; \Lfir (\Watt) $$
\citep{K98b}, where $L = 4 \pi D^{2} S$ and $D$ is the distance to
NGC~6946.  
We derive the \sdsfrfir\ map from the $S_{fir}$ map using:
$$ \sdsfrfir (\sdsfrpcu) = 5.38 \times 10^{-3}\; 
   S_{fir}\; (\Wms)\; cos\; i $$
where $cos (i = 40\arcdeg)$ gives us face-on surface values.
For the non-thermal continuum at 21 cm we use
$$ \sfrntc \: (\sfru) = 1.9 \times 10^{-22} \; 
   \nu^{0.8}\:({\mbox{GHz}})\;  \Lntc (\Whz) $$
to get the star formation rate \citep{C92} and use:
 $$ \sdsfrntc (\sdsfrpcu) = 2.30 \times 10^{-7}\; 
    S_{1.42\;Ghz}\; (\jyb)\; cos\; i $$ 
to build the \sdsfrntc\ map.  
The \sdsfr\ maps were azimuthally averaged in 30\arcsec\ radius
bins.

The results are shown in Figure \ref{sfr}a.  
The SFRs derived from the FIR and radio continuum differ by a 
factor of two. 
\citet{B03} finds that the non-thermal continuum from low 
luminosity galaxies is substantially suppressed, and so
 calibrations of the SFR from radio continuum from surveys that do 
not take into account the overall luminosities of the constituent
galaxies tend to underestimate the SFR. Bell provides an alternate
calibration:
$$ \sfrntc \: (\sfru) = \frac{5.5 \times 10^{-22}}
      {0.1 + 0.9(L/L_c)^{0.3}} \; \Lntc \: (\Whz) $$
where the total galactic flux satisfies the condition, 
$L \leq L_c$, $L_c = 6.4 \times 10^{21}\; \Whz$.
In the case of NGC~6946, $L = 5.2 \times 10^{21}\; \Whz$.
This calibration of \sfrntc\ is also shown in
Figure \ref{sfr}a and agrees with the FIR derived rate.
Although the agreement with the \sfrfir\ result is clearly promising, 
this alternate calibration was established 
using total galaxy fluxes and has not been studied with respect to 
the varying non-thermal continuum fluxes within galaxies.
Figure \ref{sfr}a also shows the SFR predicted by the \citet{K98a} 
calibration of the Schmidt law using our \sdg\ map;  
in NGC~6946, it overestimates the SFR relative 
the \sfrfir\ and \sfrntc\ values.

Another way to look at the Schmidt law is in terms of the star 
formation efficiency, SFE. 
In Figure \ref{sfr}b we express the SFE for each of our SFRs
as \Tsfe, the time it would take to
consume all the gas at the given star formation rate, \Tsfe~$ = \sdsfr/\sdg$. 
Using the \sfrfir\ and Bell's calibration for \sfrntc, the mean \Tsfe\ 
for NGC~6946 is $2.8\pm0.8$ Gyr and $\sdsfr \propto \sdg$.
This is consistent with the mean \Tsfe\ \ of 2.6 Gyr
for undisturbed galaxies found by \citet{RY99}.
The $N = 1.4$ Schmidt law is also shown on Figure \ref{sfr}b; 
the Schmidt law over-predicts the SFE and its dependence on \sdg.
\citet{WB02} found $N$ values as low as 1.1 for the galaxies
in their sample.
\citet[][see references within]{E02} has noted that the Schmidt 
law is inconsistent with observations that the SFE is roughly 
constant across range of galactic environments.  
This certainly is the case for NGC~6946.

On kpc scales, the SFE for the nucleus of
NGC~6946 does not differ from that of the disk; in other words,
the strong central FIR emission from the nucleus NGC~6946 
results from having more molecular material from which to form 
stars rather than any increase in the star formation efficiency.
This can be seen in Figure \ref{rsf} where we show the
radial distribution of the ratio of \sdg\ to 21 cm continuum;
the ratio does not decline at the nucleus.
The lowest values (highest efficiencies) are found between 2\arcmin\
and 4\arcmin\ radii (3.5 to 7 kpc).
The highest values (lowest efficiencies) are at the largest radii.
In two other late type galaxies for which we have a similar data
set, IC~342 and M83, there is a noticeable drop in log(\sdg/\ftocm) 
from the peak values in the disk to the nucleus; an indication of 
a elevated star formation efficiency relative to the disk rates
\citep{CTHLMH01, CTBHM02}.
Using the definition of \citet{RY99}, NGC~6946 would not be
a starburst galaxy because it does not have a gas cycling
time shorter than $10^9$ years.
Other authors have come to a similar conclusion about the
star formation efficiency in NGC~6946: 
\citet{MT04} based on a determination of the nuclear molecular mass from
high resolution observations of CO isotopes and multiple line 
transitions, and \citet{Mea93} by evaluating the SFR using CII maps. 

\subsection{Is There an Edge to the CO Disk in NGC~6946?}

Does the outer edge of our CO(1-0) map represent a real threshold
for the formation of molecular clouds? 
Clearly in NGC~6946, there is not a steep or marked decline in the 
gas as we detected in M83; that galaxy has a real gas edge, in
both CO and HI, possibly due to tidal interaction \citep{CTBHM02}.
In NGC~6946 the exponential disk continues as far as we can detect
it (Figure \ref{sdg_r}). But there may be other subtle changes in the 
gas with radius and we examine these below.

We use our Galaxy as guide on what to expect for the radial 
falloff in gas and star formation in a spiral galaxy.  
The number of HII regions per kpc$^2$, \sdmh,
and \sdhi\ at  $R_{\sun}$ and 12 kpc (rescaled to $R_{\sun}$ = 8 kpc)
for the Milky Way from \citet{WBH88} are presented in Table 4. 
In the Galaxy, between 8 and 12 kpc 
$N_{HII}$/kpc$^{2}$ has fallen off by a factor of 15, \sdmh\ by a 
factor of 20, while \sdhi\ stays relatively flat.\footnote{Caveat: these 
are not azimuthal averages, but rather
represent volumes at the sun and a variety of viewing angles
centered on a Galactic anticenter vector, and may not be 
representative of the Galaxy as a whole.}
Star formation per unit \mh\ remains relatively constant over this 
range of radii.
\citet{WBH88} found evidence for molecular outflows at distances
as large as 16 kpc.
\citet{LAG93} detected Galactic molecular gas in a search for CO absorption
lines against continuum sources; their findings suggest cold molecular gas
may be far more abundant at large radii than previously suspected; they
suggest the molecular gas mass may be 4 times the HI mass at 12 kpc.
\citet{MK88} detect a sparse distribution of Milky Way molecular 
clouds at 13 kpc with a mean $\Trs \sim 3$ K and a mean radius of 20 pc.  

The edge of the observed CO disk in NGC~6946 lies midway between
8 and 12 kpc for an assumed distance of 6 Mpc.
Comparing NGC~6946 to the Milky Way (Table 4): at 8 kpc (4.6\arcmin\ )
$N_{HII}$/kpc$^{2}$ is a factor of four less in NGC~6946 than 
the Galactic value at $R_{\sun}$, and could be due to extinction in \ha. 
\sdmh\ is the same for both galaxies (within the errors), 
and \sdhi\ is at least twice the Galactic value.
At 12 kpc, $N_{HII}$/kpc$^{2}$ is the same as Galactic
(extrapolating 20\arcsec\ in 
radius past the radial limit of the \citet{HK83} data set), 
\sdmh\ is no longer detected in NGC~6946 by our observations, and 
\sdhi\ is at least as large as the value for the Milky Way.
We conclude that given the similarity of $N_{HII}$/kpc$^{2}$
and \sdhi\ between these two galaxies, we would expect
that \sdmh\ in NGC~6946 should be also be $>0.1$ \sdu\ at 12 kpc.

Complicating these comparisons is the uncertainty in the NGC~6946 distance.
To date, there are no Cepheid-based distance estimates to NGC~6946,
which has a low Galactic latitude.
Recent estimates are (in Mpc): $5.7\pm0.7$ based on type II SN
\citep{ESK96}, $6.4\pm0.4$ based on the brightest stars \citep{SKT97},
and $5.9\pm0.4$ based on blue supergiants \citep{KSH00}.
As with any comparison based on absolute scales, our comparison
changes if the the 6 Mpc distance estimate is
incorrect by more than $\pm0.5$ Mpc.

The mapping of the CO disk presented here is sensitivity-limited; 
our 3$\sigma$ detection limit is 2-3 \sdu. 
A combination of low \Tex\ and a low filling factor for molecular 
clouds in the beam combine to create a detection limit rather 
than a real threshold. 
There is also the possibility that declining metallicity
in the outer disk lowers the CO emissivity relative to \Nmh,
although this effect has yet to be seen.
There are probably clouds here, but more widely 
dispersed than they are in the inner parts of the NGC~6946.

But our suspicion that there is substantial molecular gas at large
galactic radii is not based entirely on a comparison to the Milky Way.
There are theoretical reasons to expect molecular gas at radii 
beyond our detected CO edge \citep{E89,E93,EP94,HSA95}.  
According to \citet{EP94} molecular clouds are not formed where 
the ISM pressure, 
\Pism, becomes too low to facilitate the formation of cold, dense clouds.  
Or conversely, molecular clouds form in regions where $\Pism > \Pmin$.
The outer HI disk of NGC~6946 has considerable surface density 
structure that exists on kpc in-plane size scales and where \Pism\
may locally exceed this minimum.
To calculate the mid-plane gas pressure we use:
$$ \Pism = 0.5\; \pi\; G\; \sdg^2 $$
where we have ignored the stellar contribution to the pressure
which can only act to increase the value \citep{EP94}.  
Along the outer (R$\sim 6\arcmin$, 10 kpc) HI gas arms, 
$\Pism \sim 2 \times 10^{3}$ \pu\ which is close to 
\Pism\ values at the edge of the CO disk.  
The outer gas arms are where one would expect to find molecular gas,
if we had deeper maps.
Several of the HII regions identified by \citet{HK83} are located 
beyond the edge of our CO disk, particularly to the south, and 
it is likely that they are associated with molecular clouds that are too
beam-diluted and cold to show up in our images.
\citet{FWGH98}  have detected \ha\ emission associated with the formation 
of massive stars, up to 3\arcmin\ beyond our CO disk edge and precisely at 
the local maxima in the outer HI arms of NGC~6946.  
These star forming outer arms would be good places to look for
CO with a smaller beam and higher sensitivity.

Based on theoretical arguments and the detection of \ha\ in NGC~6946 
at large galactic radii, the answer is: no, the ``edge" of the CO disk
at a galactocentric radius of 5\arcmin\ ($\sim$ 9-10 kpc) in our images
is not a real edge, it is a detection limit.
There is no significant change in properties to indicate that the 
disk is undergoing a change at this radius.
There could easily be molecular clouds that escape detection beyond 
this radius

\section{Conclusions}

\noindent We have obtained deep CO(1-0) and CO(2-1) images of 
a 10\arcmin~$\times$~10\arcmin\ region centered on the Sc galaxy, 
NGC~6946, with
55\arcsec\ and 27\arcsec\ resolution using the NRAO 12 Meter Telescope. 
We have combined these deep CO images with VLA HI images to obtain 
images of the neutral gas in this galaxy.
To summarize our findings: \\
1) The CO(1-0) and CO(2-1) disks shows many of the same features seen 
in the optical disk: a strong nuclear peak of emission, molecular gas
arms spatially coincident with the asymmetric optical arm
pattern, and a 10\arcmin\ diameter inner disk filled with CO emission.
The CO disk is roughly exponential with a scale length of 
$1.2\arcmin \pm 0.2\arcmin$ or $2 \pm 0.3$ \Dcor\ kpc.\\
2) We obtain $3.3 \times 10^{9}$ \Dcorsqr\ \Mo\ for the molecular 
hydrogen gas mass of NGC~6946. 
This can be compared to the total atomic hydrogen gas mass of 
$\Mhi = 7.0 \times 10^{9}$ \Dcorsqr\ \Mo\ (which includes a VLA missing
flux estimate). 
Molecular hydrogen constitutes 1/3 of the interstellar hydrogen gas mass 
and 2\% of the dynamical mass of NGC~6946.
At the nucleus, \sdmh\ = 170 \sdu, 
the molecular gas surface density is $> 20$ \sdu\ along the strong 
optical arm pattern, 
and $\sim 10$ \sdu\ regions beyond.  \\
3) The mean value for $\Rco =I_{CO(2-1)}/I_{CO(1-0)}$ in the 
the nucleus and optical arm pattern  is $\sim 0.83\pm0.14$.  
This is consistent with optically thick gas, with an excitation 
temperature of 10-15 K.
Interarm CO shows the largest variation in \Rco, with a range of 
0.35 to 2.
Only 4\% of the CO ratio map has $\Rco < 0.6$ which is likely to be
cold or subthermally excited CO ($\Tex < 4$ K). 
Fully half of the CO disk of NGC~6946 has $\Rco > 1$. 
The high \Rco\ appears to indicate the presence of widespread,
optically-thin gas in between the spiral arms. \\
4) Molecular gas dominates the total gas surface density within a 
galactocentric radius of 3.5\arcmin\ (6 kpc).
CO emission fills the central depression in the HI disk. 
Beyond the nuclear region, there is a good correlation between
the CO and HI gas, which both peak on the spiral arms.
The inner disk arm pattern transitions smoothly from predominantly 
CO arms to predominantly HI arms at the outskirts of the CO disk.
At the nucleus, \sdg\ = 240 \sdu , the total gas surface density is
$> 20-45$ \sdu\ along the optical arm pattern,
and $\sim 5-10$ \sdu\ along the outer gas arms 
($\sdg = 1.36 (\sdmh + \sdhi)$).
CO(2-1) emission peaks are coincident with clumps of HI, \ha, and
FIR emission outside the nuclear region, presumably an indication of 
warmer molecular gas and dissociated \mh. \\
5) We find strong correlations between star formation 
tracers (except for \ha) and CO emission in NGC~6946.  
The correlation coefficients are $\sim 0.95$.
When \sdg\ is used instead of just the molecular gas 
tracer in a radial comparison, the dependence on galactic radius is reduced. 
Star formation is more closely related to the total gas surface density
than the molecular gas surface density alone, a Schmidt law relationship.
This also implies star formation is likely well beyond the radial limits
of our observed CO disk in regions of predominantly HI gas.
Surprisingly, \ha\ is not as well correlated with CO or \sdg\
as are the other star formation tracers.
The correlation coefficient is significantly less, $\sim 0.65$, and
the ratio formed from \ha\ and CO or \sdg\ declines more rapidly with
galactic radius.\\
6) The star formation efficiency in the disk of NGC~6946 is relatively
uniform; 2.8 Gyr when expressed in terms of the gas consumption
time-scale.  This uniformity implies $\sdsfr \propto \sdg$.  
The star formation rate at the nucleus is due to the high molecular 
surface density rather than an elevated star formation efficiency. \\
7) The edge of the observed CO disk represents a detection limit 
rather than a threshold for molecular cloud formation. 
We suspect the exponential fall of the CO disk continues
smoothly past our detection limit.
The mid-plane gas pressure in the outer HI arm structure,  
$\Pism \sim 2 \times 10^{3}$ \pu\, is close to the value at the
limits of the observed CO disk.  
These gas arms should support the formation of molecular clouds
which could be detectable in higher resolution observations.

\acknowledgments
The authors would like to thank the NRAO 12m telescope operators, 
Duane Clark, Paul Hart, Victor Gasho and Harry Stahl, 
for their help and company during the many observing sessions in 
which the data for NGC~6946 and other galaxies were acquired.
We thank L. Tacconi giving for us permission to retrieve and 
use the HI 21 cm line observations from the VLA archive.
And we appreciate the careful reading and critical suggestions
made by the anonymous referee. 
This work was supported in part by NSF grants AST00-71276 and
AST03-07950.
We made use of the NASA/IPAC/IRAS HiRes data reduction
facilities as well as STScI Digital Sky Survey facilities.
The Digitized Sky Surveys were produced at the Space Telescope 
Science Institute under U.S. Government grant NAG W-2166. 

{}

\setcounter{figure}{0}

\clearpage
\begin{figure} \plotone{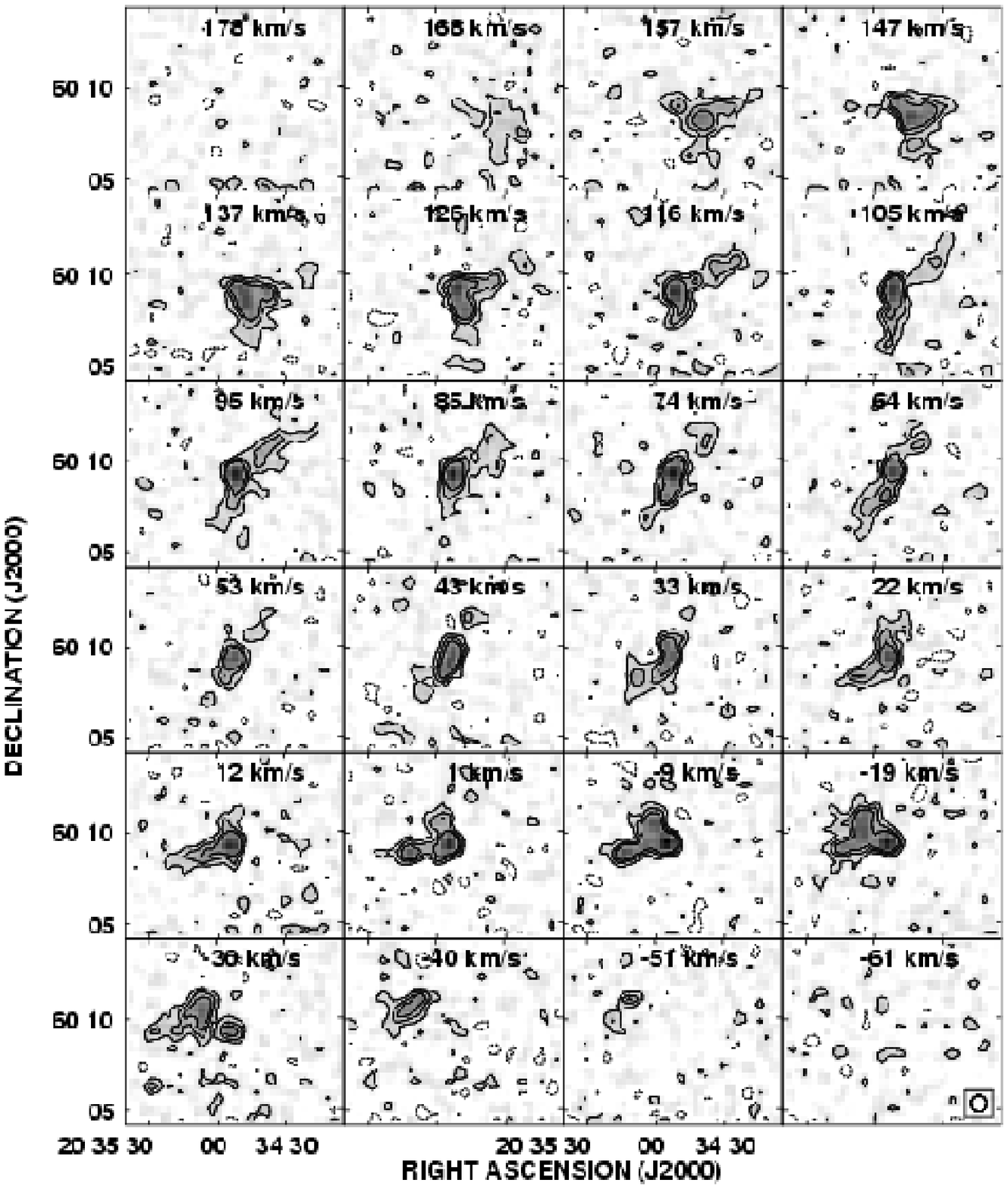}
\caption { \fcs
CO(1-0) channel maps for NGC~6946.
Grey scale ranges from 0 to 0.54 K.  Contours are at -0.070, 0.070, 0.14
and 0.21 K.  The rms noise in a channel with no emission is 0.033 K.
The 55\arcsec\ (FWHM) circular beam is displayed in the -61 \vu\ channel 
at the lower right.
Each channel is labeled with the channel velocity (LSR).
The channels are separated by 10.4 \vu.
\label{co10.chnl} }
\end{figure}
 
\clearpage 
\begin{figure} \epsscale{0.50} \plotone{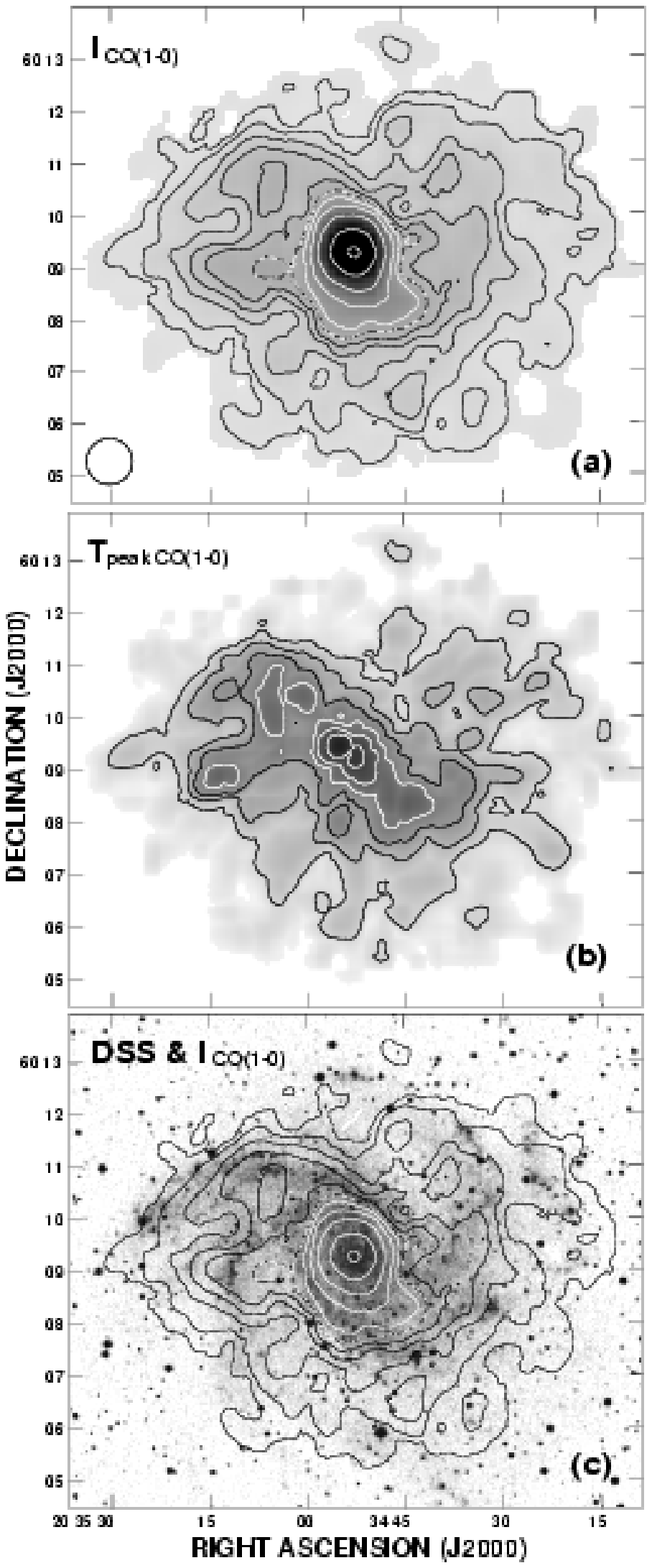} \epsscale{1.0}
\caption {\fcs
CO(1-0) integrated and peak intensity maps for NGC~6946.
a) CO(1-0) integrated intensity map.  Grey scale ranges from 0 to 50
\iu.  The contour levels are 1, 2.5, 5, 7.5, 10, 12.5, 15, 20, 30,
50, and 70 \iu.
The 55\arcsec\ (FWHM) circular beam is displayed at the lower left.
b) CO(1-0) peak intensity map.  Grey scale ranges from 0 to 0.52 K.  
The contour levels are 0.10, 0.17, 0.23, 0.30, 0.37 and 0.43 K.
c) DSS uncalibrated optical image with the same \Ioz\ contours
shown in (a).
\label{co10.m0tb} }
\end{figure}

\clearpage 
\begin{figure} \plotone{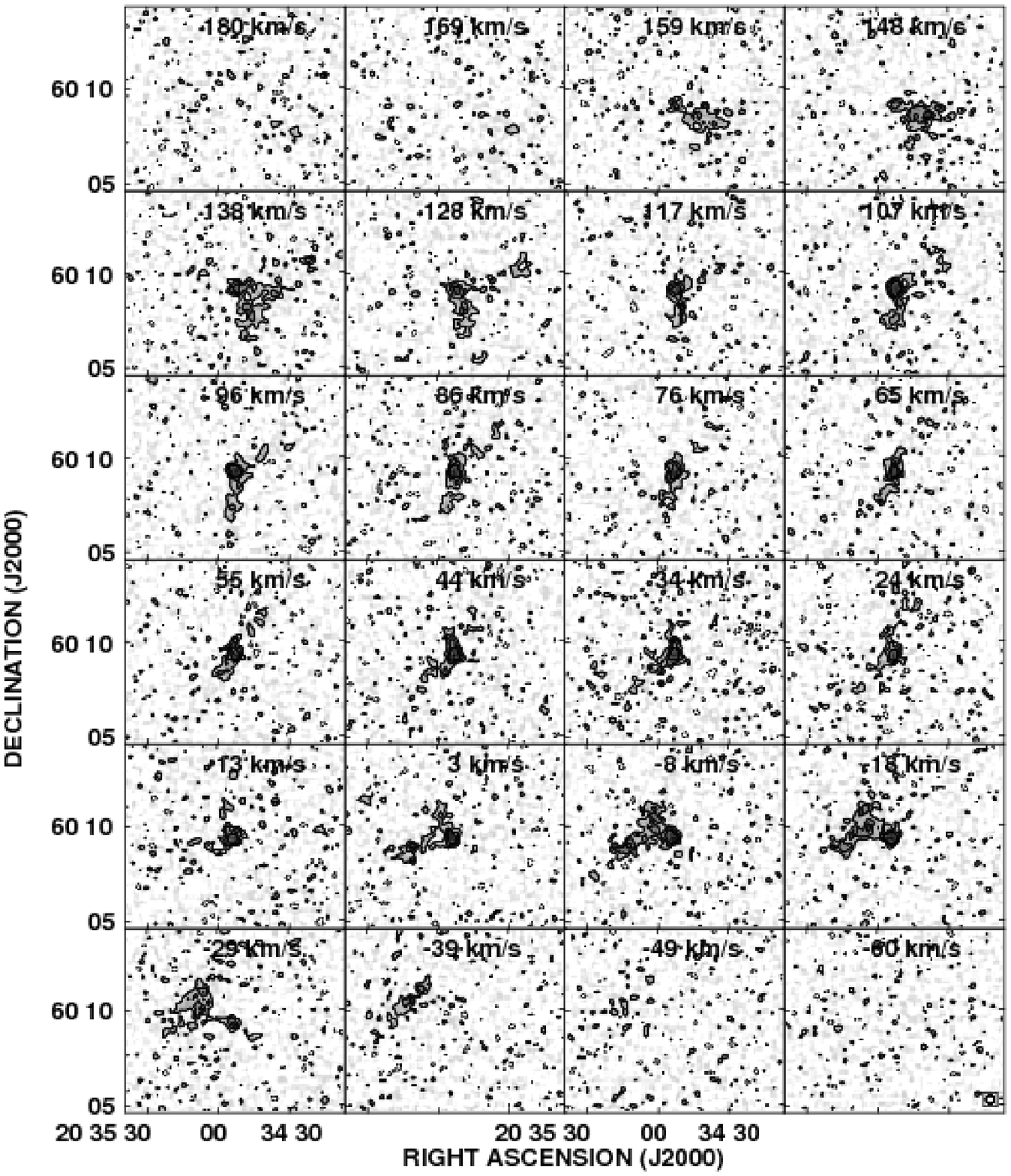}
\caption {\fcs
CO(2-1) channel maps for NGC~6946.
Grey scale ranges from 0 to 0.85 K.  Contours are at -0.17, 0.17, 0.34
and 0.51 K.  The rms noise in a channel with no emission is 0.085 K.
The 27\arcsec\ (FWHM) circular beam is displayed in the -60 \vu\ channel 
at the lower right.
Each channel is labeled with the channel velocity (LSR).
The channels are separated by 10.4 \vu.
\label{co21.chnl} }
\end{figure}
 
\clearpage 
\begin{figure} \epsscale{0.50} \plotone{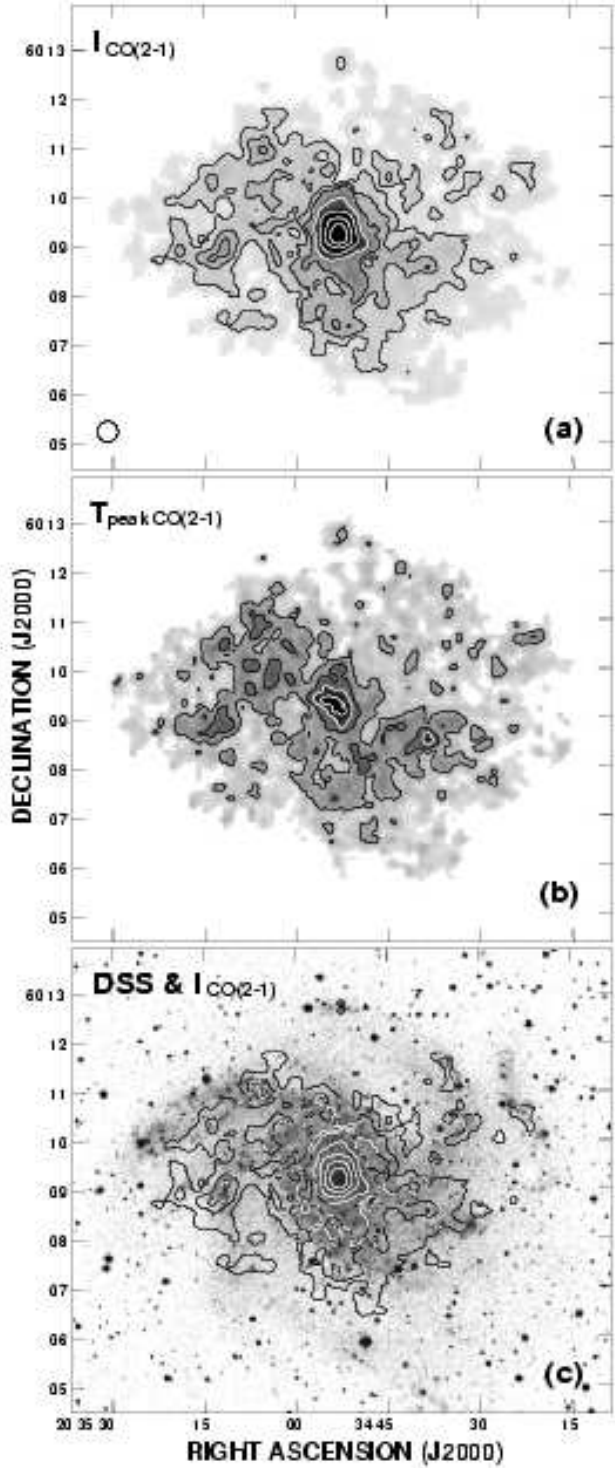} \epsscale{1.0}
\caption {\fcs
CO(2-1) integrated and peak intensity maps for NGC~6946.
a) CO(2-1) integrated intensity map.  Grey scale ranges from 0 to 50
\iu.  The contour levels are 5, 10, 15, 30, 50, 75, and 100 \iu.
The 27\arcsec\ (FWHM) circular beam is displayed at the lower left.
b) CO(2-1) peak intensity map.  Grey scale ranges from 0 to 0.8 K.  
The contour levels are 0.26, 0.43, 0.60 and 0.77 K.
c) DSS uncalibrated optical image with the same \Ito\ contours
shown in (a).
\label{co21.m0tb} }
\end{figure}

\clearpage 
\begin{figure} \plottwo{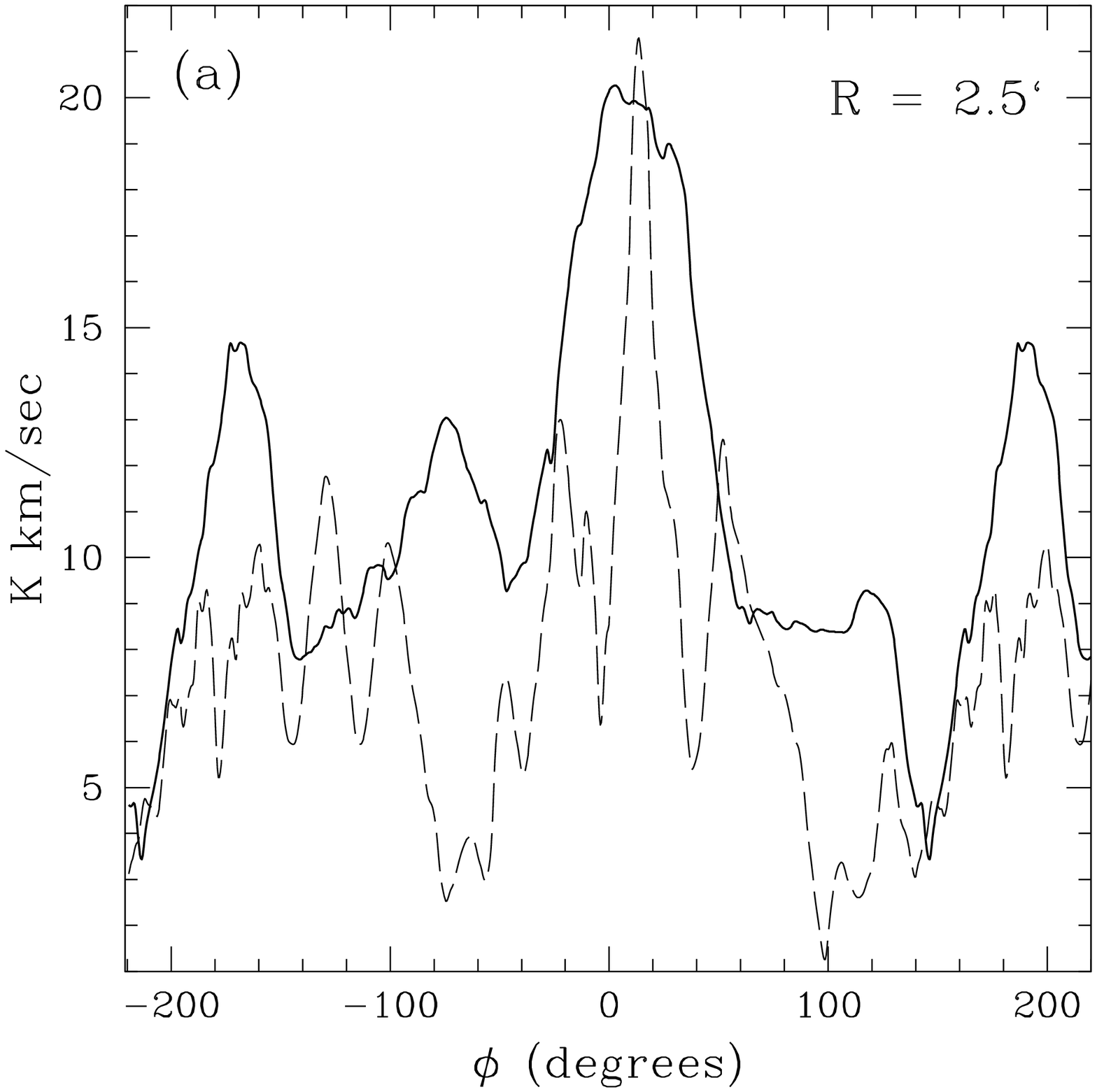}{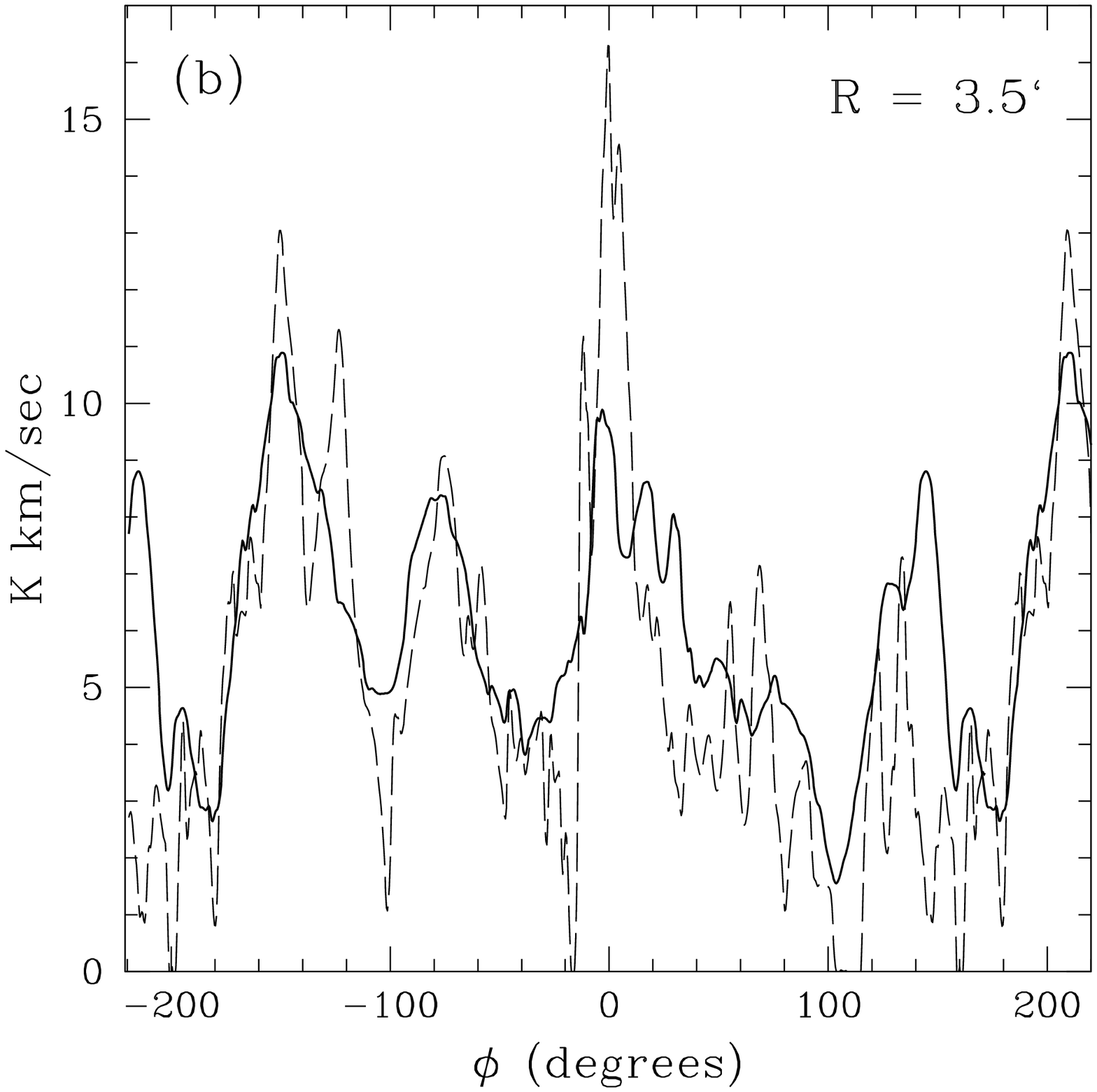}
\caption {\fcs
Azimuthal plots of CO in NGC~6946 at selected radii.
The \Ico\ maps were transformed to face-on images and the
azimuthal variation in \Ico\ was measured at constant galactic radii. 
Due south is at $\phi = 0\arcdeg$, west is at $\phi = 90\arcdeg$.
CO(2-1) is traced by the dashed line, CO(1-0) by the solid line.
The data are presented at their native beam sizes, 27\arcsec\
and 55\arcsec, for the CO(2-1) and CO(1-0) respectively.
a) The azimuthal variation at R = 2.5\arcmin\  (4.4 kpc).
b) The azimuthal variation at R = 3.5\arcmin\  (6.1 kpc).
\label{coaz} }
\end{figure}

\clearpage 
\begin{figure} \plotone{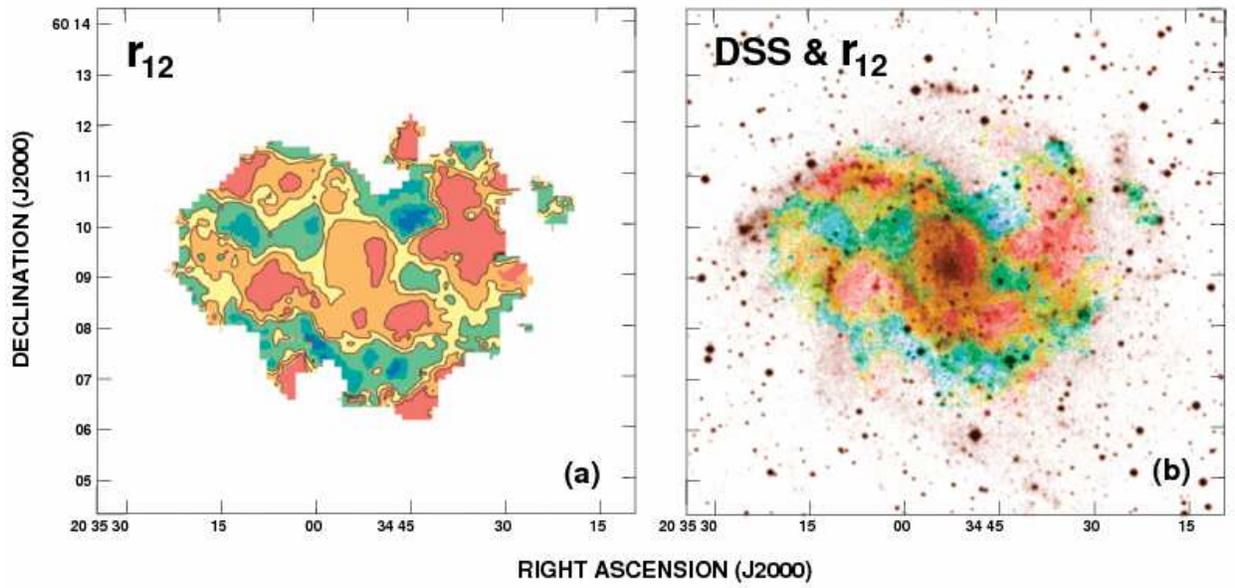}
\caption {\fcs
The CO(2-1)/CO(1-0) ratio in NGC~6946.
a) $\Rco\; = \Ito / \Ioz$.  The contours are at \Rco = 0.8, 1.0 and 1.2.
For the colored regions: red represents $\Rco < 0.8$; orange for
$0.8 > \Rco > 1.0$; yellow for $1.0 > \Rco > 1.2$; green for $\Rco > 1.2$;
blue for $\Rco \sim 2$.
b) Digital Sky Survey optical image with \Rco\ in color.  
The color scheme roughly follows that of figure a) and is intended to
show the relationship between the regions of high/low \Rco\ and
the optical stellar disk.
\label{Rco} }
\end{figure}

\clearpage 
\begin{figure} \plotone{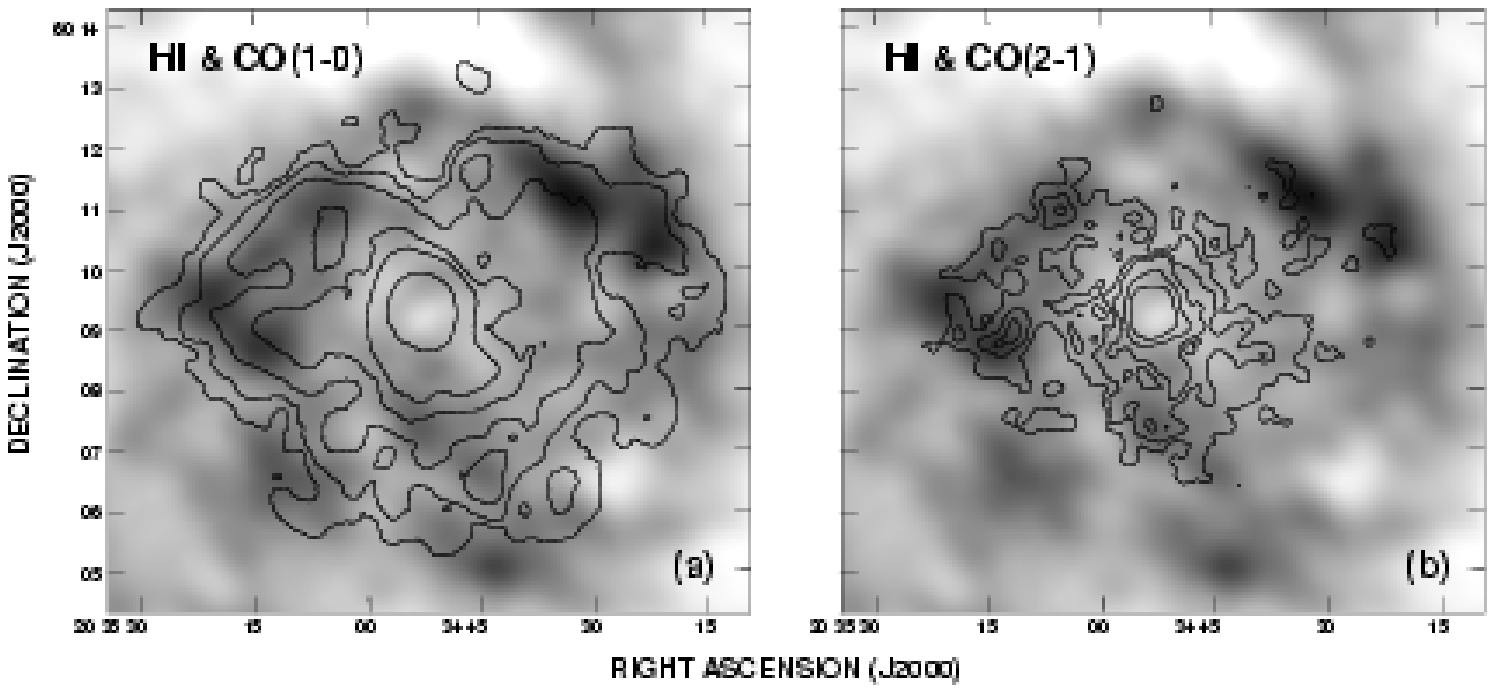}
\caption {\fcs
CO and HI in NGC~6946.
a) \Ihi\ (grey) and \Ioz\ (contours).
Grey scale ranges from 0.5 to 3.7 \su.
The  \Ioz\ contours and are at 1, 2.5, 5, 10, 15 and 35 \iu.
b) \Ihi\ (grey) and \Ito\ (contours).
Grey scale ranges from 0 to 3.7 \su.
The  \Ito\ contours and are at 5, 10, 15, 30 and 50 \iu.
\label{hico} }
\end{figure}

\clearpage 
\begin{figure} \plotone{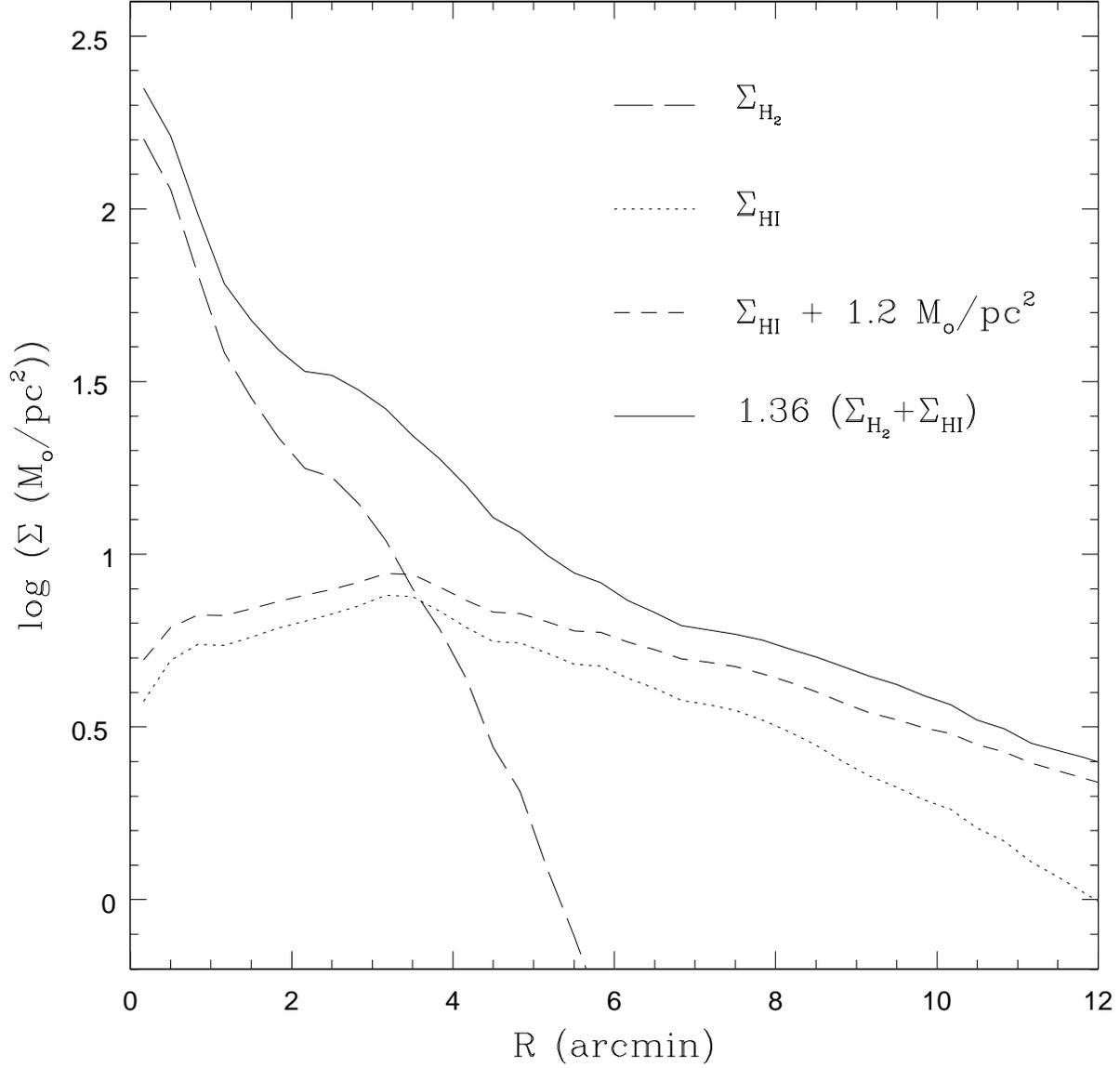}
\caption {\fcs
Gas surface density vs radius in NGC~6946.
Plotted are: \sdmh\ derived using the standard conversion factor;
\sdhi\ from the observed emission;
\sdhi\ increased for the VLA missing flux estimate;
and the total gas surface density increased by a factor of 1.36 to
include the He and heavier element content.
All have been corrected for inclination, (cos i).
At our assumed distance of 6 Mpc, 1\arcmin\ = 1.75 kpc.
\label{sdg_r} }
\end{figure}

\clearpage 
\begin{figure} \plottwo{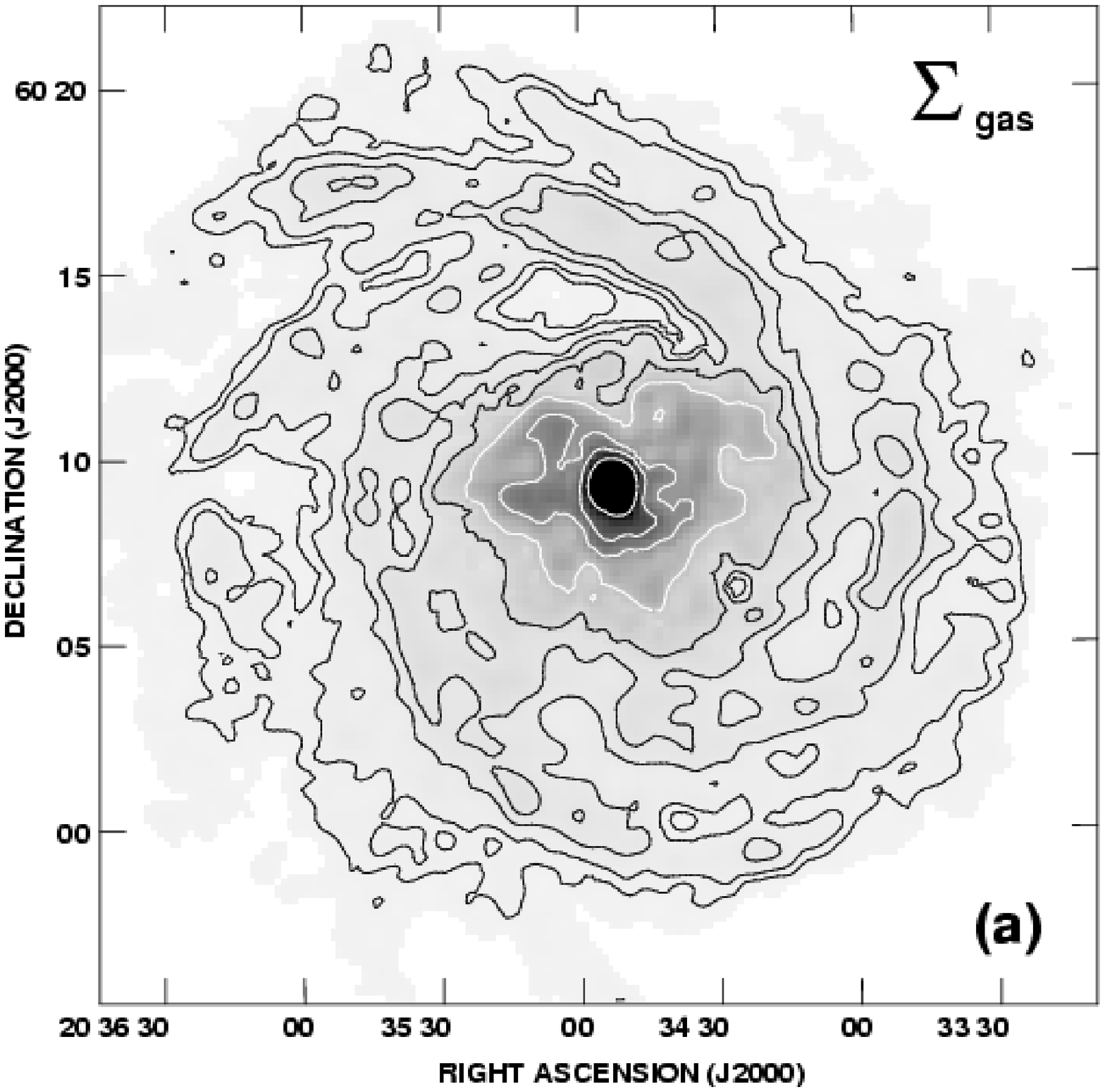}{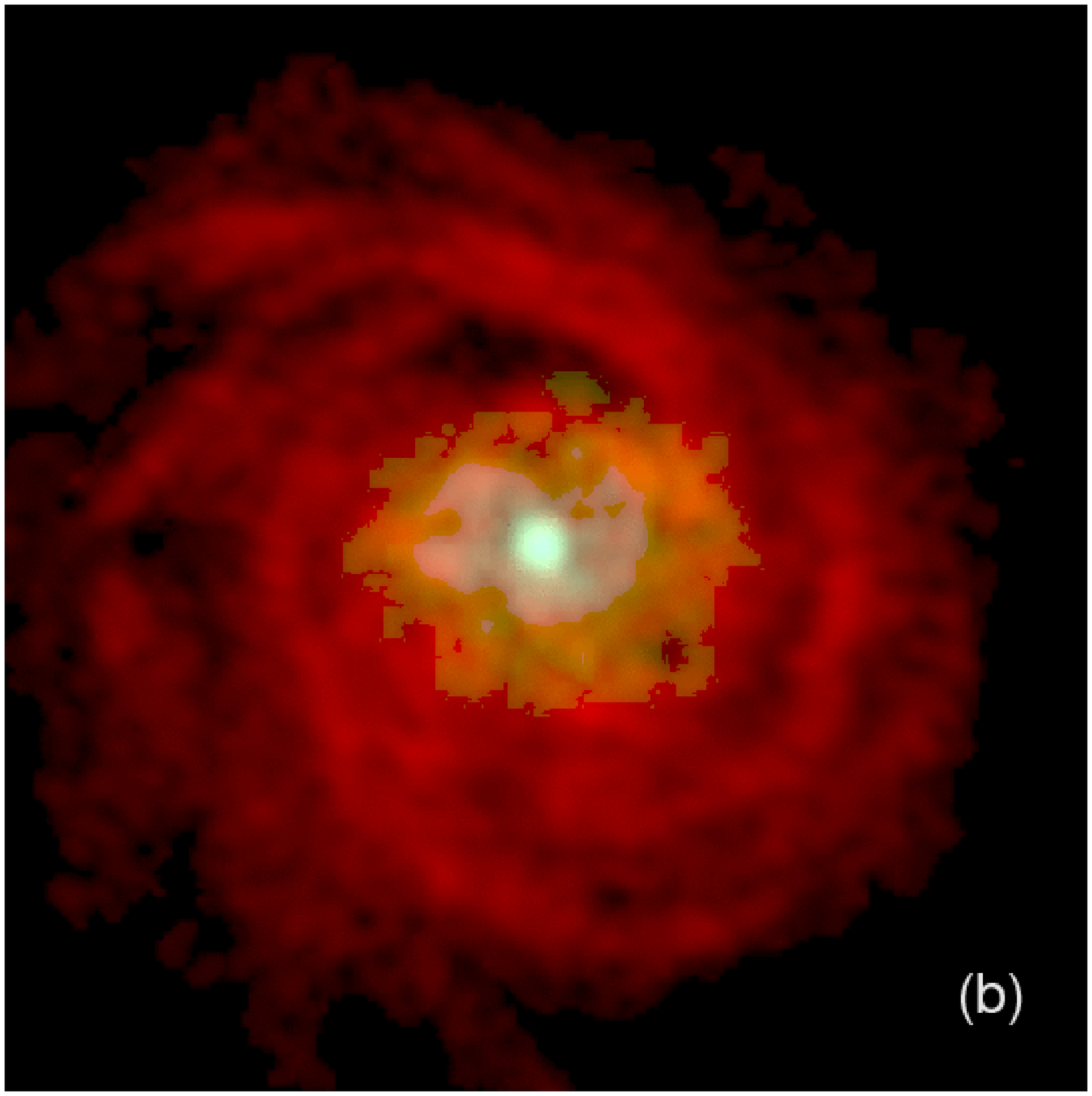}
\caption {\fcs
The neutral gas surface density in NGC~6946.
a) A map of the total neutral gas surface density, corrected
for inclination (cos(40\arcdeg)) and increased by a factor of 1.36 for 
the He and heavier element content.  
This map does not include a missing HI flux estimate.
Contours are at 1, 2, 4, 6, 10, 20, 40, 60 and 100 \sdu.
b) A false color image with CO(1-0) in green and HI in red.
Regions of nearly equivalent column densities show up in orange.
The CO(2-1) emission has been added in blue to bring out the
bright CO nucleus.
\label{sdg} }
\end{figure}

\clearpage 
\begin{figure} \plotone{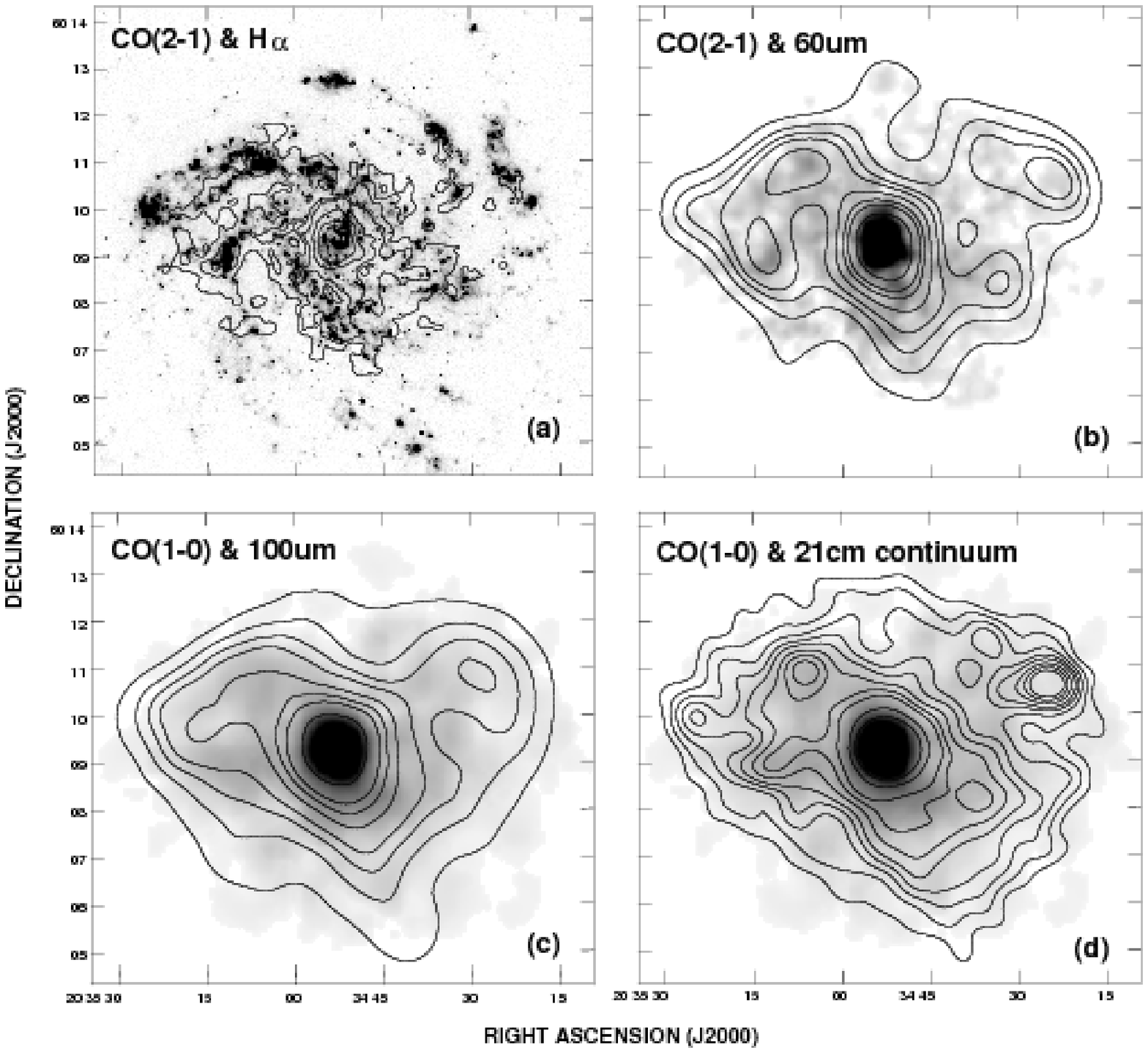}
\caption {\fcs
CO and star formation tracers in NGC~6946.
The CO(1-0) and CO(2-1) beam sizes are 55\arcsec\ and 27\arcsec\ 
respectively.
a) \Ito\ in contours and \ha\ in grey.
The \ha\ image is from \citet{FWGH98} and is not presented here
in calibrated units. 
The \ha\ image was regridded to 2\arcsec\ pixels.
The \Ito\ contours are at 5, 10, 15, 30, and 50 \iu.
b) \Ito\ in grey with IRAS 60 \fir\ contours.
The grey scale ranges from 0 to 40 \iu.
The 60 \fir\ contours are at 10, 20, 30, 40, 60, 90, 120 and 200 \Mjys.  
The 60 \fir\ beam size (FWHM) is 45\arcsec~$\times$~41\arcsec,
$pa = 21\arcdeg$\, and the rms noise level is 0.6 \Mjys.
c) \Ioz\ in grey with IRAS 100 \fir\ contours.
The grey scale ranges from 0 to 40 \iu.
The 100 \fir\ contours are at 20, 40, 60, 80, 120, 160, 200 and 300
\Mjys.  The 100 \fir\ beam size (FWHM) is 69\arcsec~$\times$~65\arcsec,
$pa = 20\arcdeg$\, and the rms noise level is 2 \Mjys.
d) \Ioz\ in grey with 21 cm continuum contours.
The grey scale ranges from 0 to 40 \iu.
The 21 cm contours are at 5, 7, 10, 12, 18, 22, 25, 36 and 48 \mjyb.
The beam size size (FWHM) is 49\arcsec~$\times$~42\arcsec,
$pa = 73\arcdeg$\, and the rms noise level is 1.2 \mjyb.
\label{cosf} }
\end{figure}

\clearpage 
\begin{figure} \plottwo{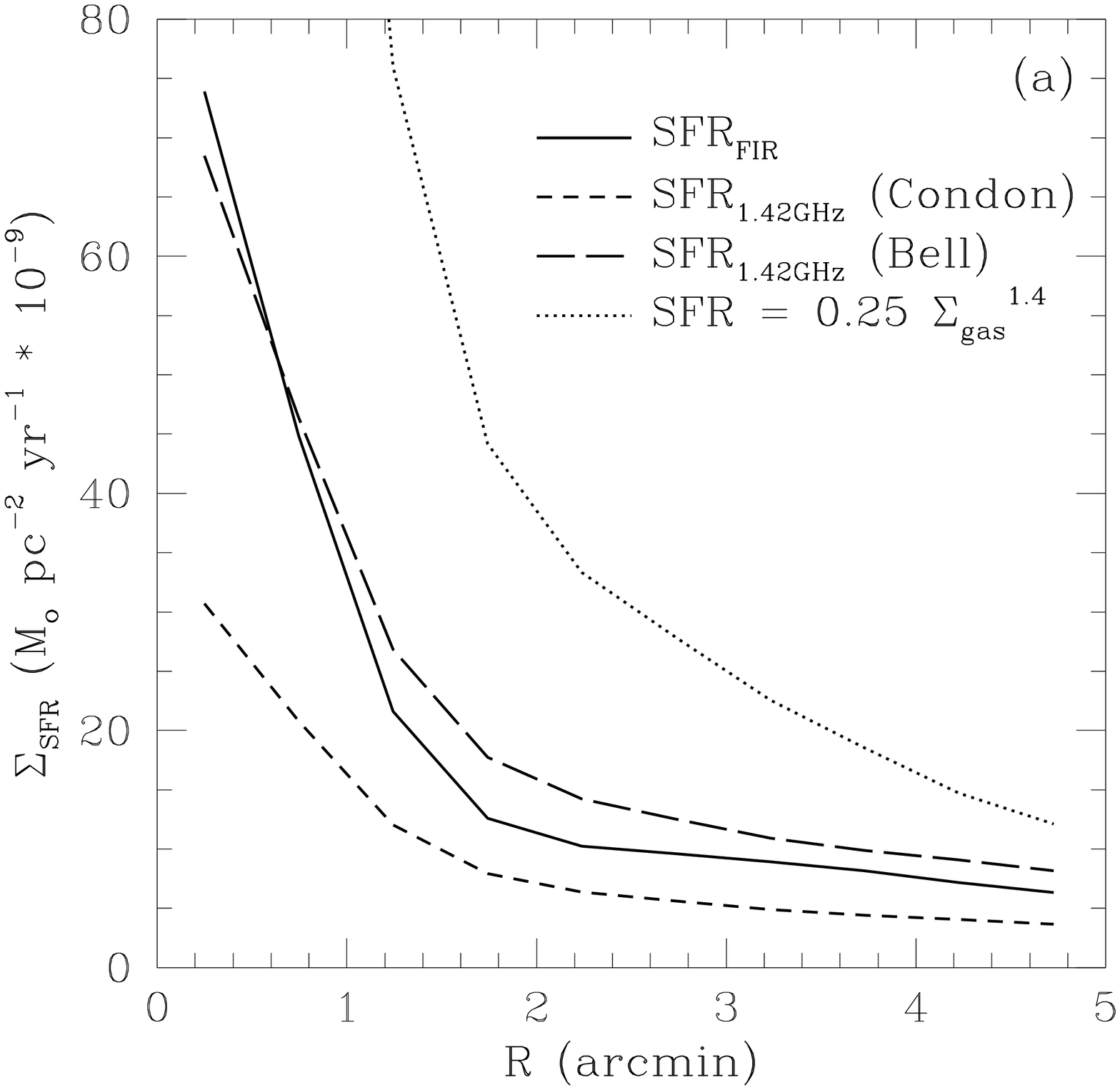}{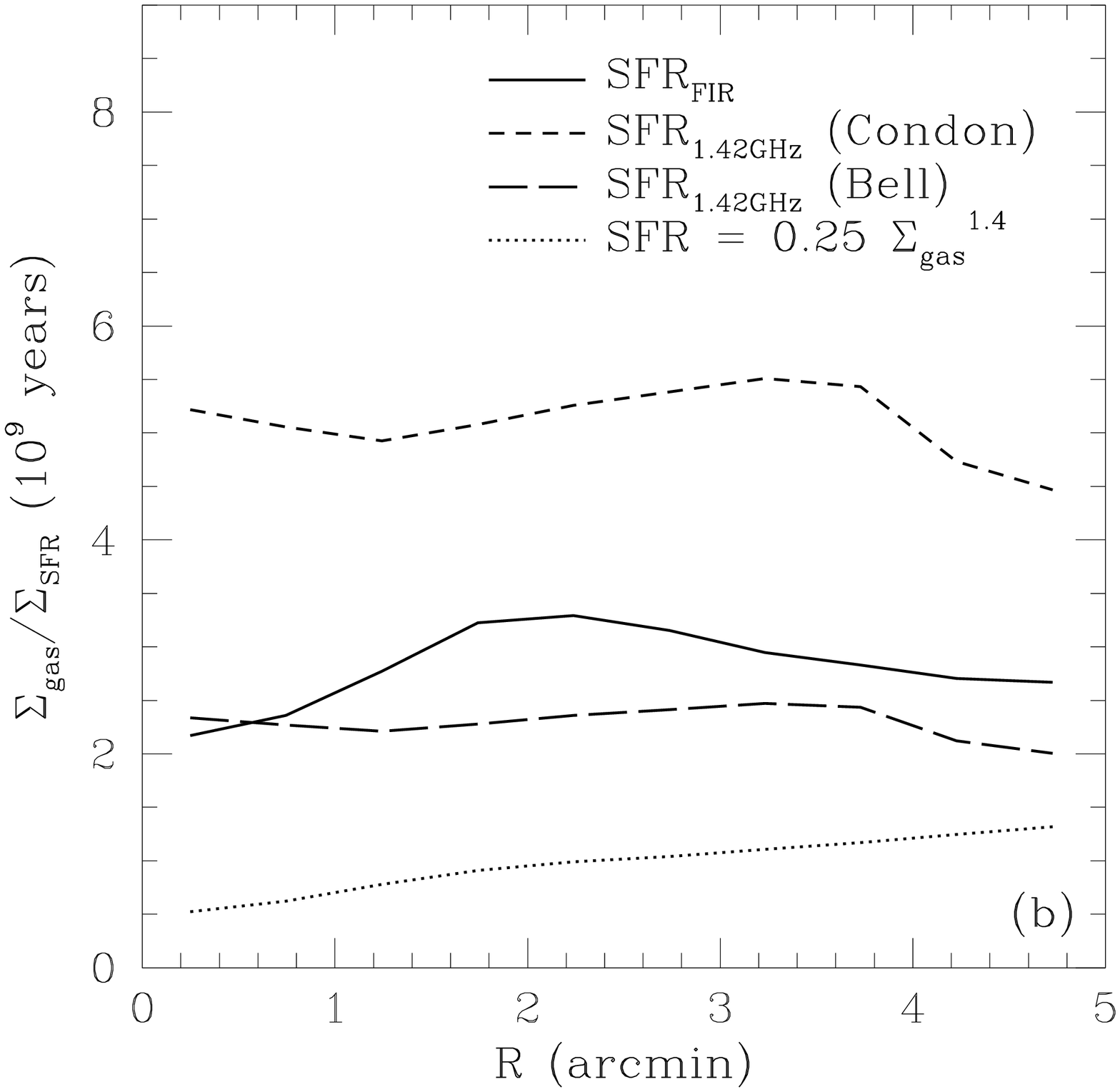}
\caption {\fcs
Radial star formation rates and efficiencies in NGC~6946.
a) \sdsfr\ derived from the IRAS luminosities, \Lfir; 
non-thermal radio continuum luminosity, \Lntc, using the SFR calibration
of \citet{C92}, and \Lntc\ using the SFR calibration from \citet{B03};
and as predicted by Schmidt law of \citet{K98a} from \sdg.
b) The star formation efficiency expressed in terms of \Tsfe, the
time it would take to consume all of the molecular gas at the
given star formation rate.  
The legends in both figures indicate the source of the SFR calibration.
At our assumed distance of 6 Mpc, 1\arcmin\ = 1.75 kpc.
\label{sfr} }
\end{figure}

\clearpage 
\begin{figure} \plotone{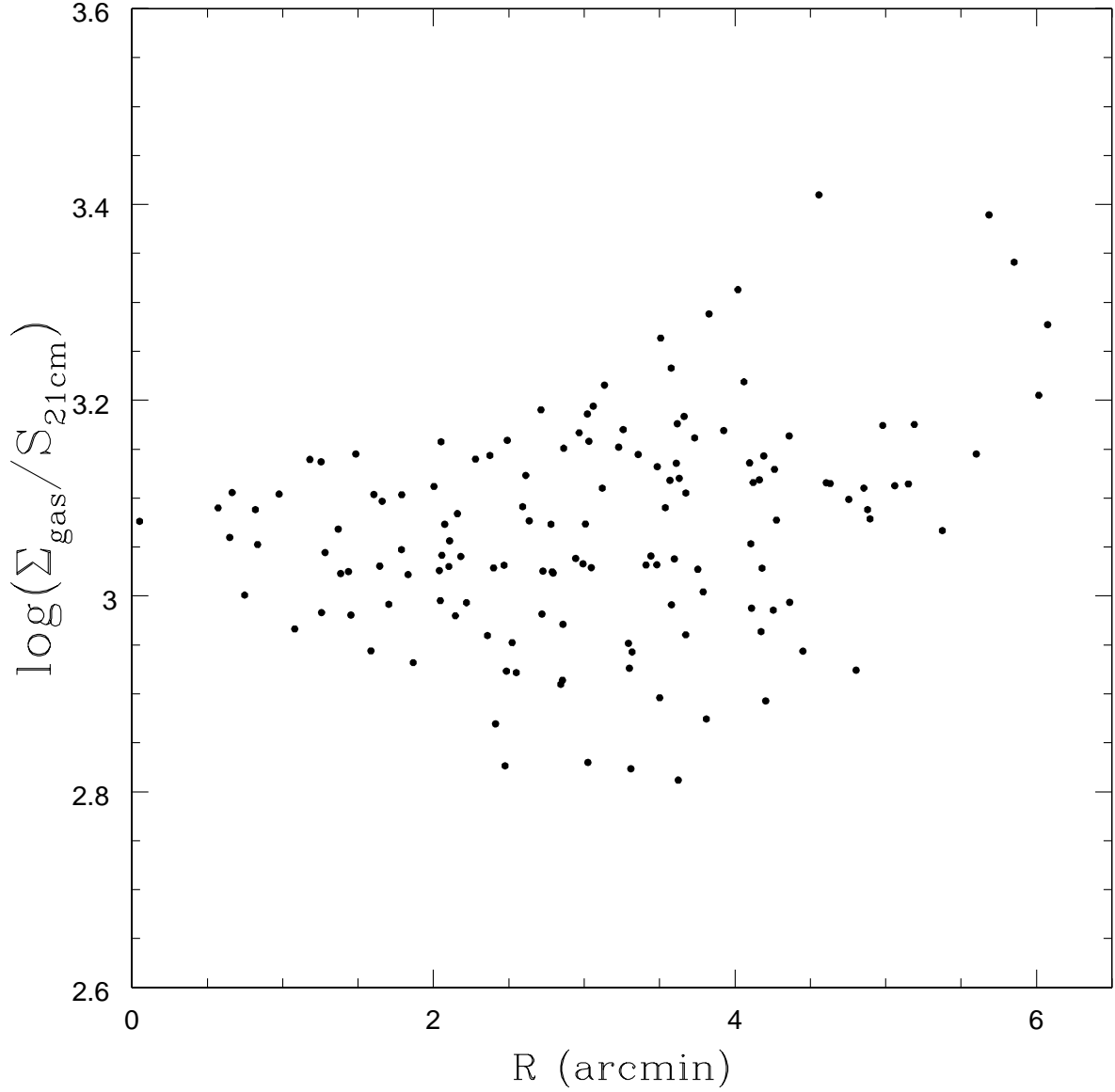}
\caption {\fcs
Ratio of total gas surface density to 21 cm continuum in NGC~6946.
The gas surface density maps was convolved to a 70\arcsec\ FWHM 
beam size and clipped at a 3$\sigma$ level prior to forming the ratios.  
The resulting ratio maps were sampled on a 35\arcsec\ grid.
At our assumed distance of 6 Mpc, 1\arcmin\ = 1.75 kpc.
\label{rsf} }
\end{figure}
 
\clearpage
\begin{deluxetable}{ll}
\tablewidth{0pt}
\tablenum{1}
\tablecaption{Global Properties of NGC 6946}
\tablehead{\colhead{Property\phm{...........................}}
          &\colhead{Value}}
\startdata

Hubble type \tablenotemark{a}          &  Scd \\
R.A. (J2000) \tablenotemark{b}         &  $20^{h}\;34^{m}\;51.\!^{s}6
                                          \; (\pm0.\!^{s}4)$ \\
Dec. (J2000) \tablenotemark{b}         &  $60\arcdeg\;9\arcmin\;8.7\arcsec
                                          \; (\pm1.6\arcsec)$ \\
$l\,^{II}, b\,^{II}$ \tablenotemark{a} &  $95.7\arcdeg, 11.7\arcdeg$ \\
Distance   \tablenotemark{c}           &  6 Mpc \\
$v_{_{LSR}}$ \tablenotemark{d}         &  50.5 $(\pm1.3)$ \vu \\
Inclination \tablenotemark{d}          &  40\arcdeg $(\pm10^{\arcdeg})$ \\
Position Angle \tablenotemark{d}       &  242\arcdeg $(\pm1^{\arcdeg})$ \\
$V_{max}$ \tablenotemark{d}            &  170 $(\pm40)$ \vu \\
$R_{V_{max}}$ \tablenotemark{d}        &  5.1\arcmin $(\pm0.7\arcmin)$ (9.0 kpc) \\
$n$ \tablenotemark{d}                  &  1.07 $(\pm0.07)$ \\
$M_{H2\; total}$ \tablenotemark{e}     &  $3.3\times10^{9}$ \Mo \\
$M_{HI\; observed}$ \tablenotemark{f}  &  $5\times10^{9}$ \Mo \\
$M_{HI}$ \tablenotemark{g}             &  $7\times10^{9}$ \Mo \\
$M_{gas}$ \tablenotemark{h}            &  $1.4\times10^{10}$ \Mo \\
$M_{dyn}$ \tablenotemark{i}            &  $1.9\times10^{11}$ \Mo \\
\phm{..................................................}    &

\enddata
\tablenotetext{a}{\footnotesize obtained from the NASA/IPAC
                  Extragalactic Database.}
\tablenotetext{b}{\footnotesize dynamical center based on fit of 
                  Brandt rotation model to the HI data.}
\tablenotetext{c}{\footnotesize Sharina et al. (1997).}
\tablenotetext{d}{\footnotesize kinematic parameters from fit of 
                  Brandt rotation model to the HI data.}
\tablenotetext{e}{\footnotesize detected, using a 
                  $2.0 \times 10^{20}$ \iu conversion factor.} 
\tablenotetext{f}{\footnotesize detected, no VLA missing flux 
                  estimate.}
\tablenotetext{g}{\footnotesize includes VLA missing flux 
                  estimate.}
\tablenotetext{h}{\footnotesize includes factor for He and heavier
                  elements: $1.36 (M_{H2} + M_{HI})$ .}
\tablenotetext{i}{\footnotesize total dynamical mass based on 
                  Brandt model fit parameters.}
\end{deluxetable}

\clearpage
\begin{deluxetable}{lcccc}
\tablewidth{0pt}
\tablenum{2}
\tablecaption{Properties of 3 Late-type Galaxies and the Milky Way}
\tablehead{\colhead{Galaxy\phm{......................}}
          &\colhead{IC342 \tablenotemark{a}}
          &\colhead{NGC 6946}
          &\colhead{M83 \tablenotemark{b}}
          &\colhead{Milky Way}}
\startdata

Hubble type    & Scd & Scd & SBc & SBbc \\

Distance (Mpc) & 3.3 \tablenotemark{c}  & 6 & 4 & ... \\

D$_{H_{2}}$ (arcminutes) \tablenotemark{d} & 20 & 11 & 11 & ...  \\

D$_{H_{2}}$/D$_{HI}$  & 0.23 & 0.38 & 0.15 & ...  \\

D$_{H_{2}}$/D$_{25}$ \tablenotemark{e} & 1.1 & 1.0 & 1.0 & ...  \\

M$_{dynamical}$ ($10^{11}$ \Mo) & 1.8 \tablenotemark{f} 
                                & 1.9 
                                & 0.7 
                                & $\sim 3$ \tablenotemark{g} \\

\Mmh$_{total}$ ($10^{8}$ \Mo) & 16 \tablenotemark{h,i} 
                              & 33 
                              & 23 \tablenotemark{i}
                              & 25 \tablenotemark{j}  \\

\Mhi$_{total}$ ($10^{8}$ \Mo) \tablenotemark{k} & 57 \tablenotemark{h} 
                                                & 70 
                                                & 62 
                                                & 48 \tablenotemark{l}  \\

\Mmh/\Mhi & 0.27 & 0.47 & 0.37 & 0.52  \\

M$_{gas}$/M$_{dyn}$ \tablenotemark{m} & 0.055 & 0.074 & 0.16 & 0.03  \\

\sdmh$^{nucleus}$ (\sdu) \tablenotemark{n} & 140 
                                           & 170 
                                           & 220 
                                           & 400 \tablenotemark{o} \\

\sdmh$^{disk}$ (\sdu) \tablenotemark{p} & 11 
                                        & 13 
                                        & 23 
                                        & 3.4 \tablenotemark{q} \\

\enddata
\tablenotetext{a}{\footnotesize data from Crosthwaite et al. (2000,2001).}
\tablenotetext{b}{\footnotesize data from Crosthwaite et al. (2002).}
\tablenotetext{c}{\footnotesize latest distance estimate based
                  on Cepheid variables Saha et al. (2002).}
\tablenotetext{d}{\footnotesize using the radius at which the azimuthal 
                  average $\sdmh \sim 0.5 \; \sdu$.}
\tablenotetext{e}{\footnotesize D$_{25}$ from RC2, de Vaucouleurs, 
                  de Vaucouleurs \& Corwin (1976).}
\tablenotetext{f}{\footnotesize dynamical mass from Brandt model using 3.3 Mpc
                  adjusted from 2 Mpc.}
\tablenotetext{g}{\footnotesize within 20 kpc, Nikiforov et al. (2000),
                  from rotation curve modeling.}
\tablenotetext{h}{\footnotesize adjusted to 3.3 Mpc distance from 2 Mpc.} 
\tablenotetext{i}{\footnotesize rescaled to a $2.0 \times 10^{20}$ \iu 
                  conversion factor.}
\tablenotetext{j}{\footnotesize Combes (1991).}
\tablenotetext{k}{\footnotesize includes VLA missing flux estimates.}
\tablenotetext{l}{\footnotesize Kulkarni \& Heiles (1987).}
\tablenotetext{m}{\footnotesize gas includes factor of 1.36 for He and
                  heavier elements.}
\tablenotetext{n}{\footnotesize \sdu\ in 55\arcmin\ beam centered on the nucleus.}
\tablenotetext{o}{\footnotesize R $<$ 400 pc, Scoville \& Sanders (1987).}
\tablenotetext{p}{\footnotesize mean \sdmh\ from observed CO disk 
                   excluding the nucleus.}
\tablenotetext{q}{\footnotesize 400 pc $<$ R $<$ l4 kpc, Scoville \& Sanders (1987).}
\end{deluxetable}

\headsep    0.0in
\textheight 9.0in

\clearpage
\begin{deluxetable}{ccc}
\tablewidth{0pt}
\tablenum{3}
\tablecaption{Radial Slope of Gas to Star Formation Tracer Ratios}
\tablehead{\colhead{Gas Tracer \tablenotemark{a}}
          &\colhead{SF Tracer \tablenotemark{a}}
          &\colhead{Slope (Uncertainty) \tablenotemark{b}}}
\startdata

\Ito & 100 \fir         & -0.196 (0.015) \\
\Ito &  60 \fir         & -0.189 (0.015) \\
\Ito &  21 cm continuum & -0.170 (0.014) \\
\Ito &  \ha             & -0.224 (0.015) \\
     &                  &       \\
\Ioz & 100 \fir         & -0.102 (0.008) \\
\Ioz &  60 \fir         & -0.095 (0.009) \\
\Ioz &  21 cm continuum & -0.076 (0.009) \\
\Ioz &  \ha             & -0.130 (0.014) \\
     &                  &       \\
\sdg \tablenotemark{c} & 100 \fir         & -0.002 (0.006) \\
\sdg &  60 \fir         &  0.005 (0.007) \\
\sdg &  21 cm continuum &  0.024 (0.007) \\
\sdg &  \ha             & -0.030 (0.013) \\

\enddata
\tablenotetext {a} {\footnotesize
Maps were convolved to a 70\arcsec\ circular beam size 
and clipped at 3$\sigma$.
The region containing the bright 21 cm continuum emission,
tentatively identified as a background galaxy, at 
\rah{20}\ram{34}\ras{24.9} \decd{60}\decm{10}\decs{38} (J2000)
was clipped from all the comparison maps.
}
\tablenotetext {b} {\footnotesize
Slope (ratio/arcminute) and rms uncertainty from a least squares
fit to the  ratio of gas tracer to formation tracer as a function 
of galactic radius, log(\gasSF) $= A \cdot r + B$. 
The maps were sampled on a grid of half beam width points (35\arcsec).
}
\tablenotetext {c} {\footnotesize
The molecular gas surface density derived from CO plus the atomic 
gas surface density derived from HI, increased by a factor of
1.36 to account for He and heavier elements.
}
\end{deluxetable}

\clearpage
\begin{deluxetable}{lrccrcc}
\tablewidth{0pt}
\tablenum{4}
\tablecaption{Comparison of the Milky Way and NGC~6946 at Large R}
\tablehead{ \colhead{} 
          & \colhead{Milky \tablenotemark{a,b}} 
          & \colhead{} & \colhead{} 
          & \colhead{NGC \tablenotemark{c}} 
          & \colhead{}  & \colhead{} \\
            \colhead{} 
          & \colhead{ Way} & \colhead{8 kpc} & \colhead{12 kpc} 
          & \colhead{6946} & \colhead{8 kpc}  & \colhead{12 kpc} }
\startdata

$N_{HII}$/kpc$^{2}$ & & 21  & $>1.4$   
                    & & $\sim 5$ \tablenotemark{d} 
                      & $\sim 1$ \tablenotemark{d} \\

\sdmh (\sdu)        & & 2.9 & $>0.15$  
                    & & $2.3\pm1.7$ & .. \tablenotemark{e}  \\

\sdhi (\sdu)        & & 2.9 & 2.9   
                    & & 7-9 \tablenotemark{f} & 3-5 \tablenotemark{f} \\

\enddata
\tablenotetext{a}{\footnotesize using R$_{\sun}$ = 8 kpc.}
\tablenotetext{b}{\footnotesize values from discussion in
                                Wouterloot et al. (1988)}
\tablenotetext{c}{\footnotesize using distance to NGC~6946 = 6 Mpc.}
\tablenotetext{d}{\footnotesize the HII region positions from
                                Hodge \& Kennicutt (1983) were
                                sampled on a grid of points separated by 
                                10\arcsec. 
                                The points were regrouped
                                into 20\arcsec\ wide galactic radius bins.
                                The 12 kpc value was extrapolated from
                                the resulting $N_{HII}$/kpc$^{2}$ curve.}
\tablenotetext{e}{\footnotesize beyond our measured CO disk.}
\tablenotetext{f}{\footnotesize including the estimated flux
                                missing from the VLA map
                                would increase these by 1.2 \sdu.}
\end{deluxetable}


\begin{thebibliography}{}

\setlength{\baselineskip}{0.6cm}
\setlength{\parskip}{-0.15cm}

\bibitem[Adler, Allen, \& Lo(1991)]{AAL91}
    Adler, D. S., Allen, R. J., \& Lo, K. Y. 1991, \apj, 283, 475
\bibitem[Arp(1966)]{A66} 
    Arp, H.\ 1966, \apjs, 14, 1 
\bibitem[Arsenault(1989)]{A89}
    Arsenault, R. 1989, \aap, 217, 66
\bibitem[Ball et al.(1985)]{BSSLS85}
    Ball, R., et al.  1985, \apj, 298, L21
\bibitem[Bell(2003)]{B03}
    Bell, E.F. 2003, \apj, 586, 794
\bibitem[Belly \& Roy(1992)]{BR92}
    Belley, J. \& Roy, J. 1992, \apjs, 78, 61
\bibitem[Bicay et al.(1989)]{BHC89}
    Bicay, M.D., Helou, G. \& Condon, J.J.  1989, \apj, 338, L53
\bibitem[Bicay \& Helou(1990)]{BH90}
    Bicay, M.D. \& Helou, G. 1990, \apj, 362, 59
\bibitem[Boulanger \& Viallefond(1992)]{BV92}
    Boulanger, F. \& Viallefond, F.  1992, \aap, 266, 37
\bibitem[Carignan et al.(1990)]{CCBV90}
    Carignan, C., et al.  1990, \aap, 234, 43
\bibitem[Casoli et al.(1990)]{CCVCB90}
    Casoli, F., et al.  1990, \aap, 233, 357
\bibitem[Castets et al.(1990)]{Cea90}
    Castets, A., et al. 1990, \aap, 234, 469
\bibitem[Catalogued Galaxies and Quasars in the IRAS Survey(1985)]{CGQ85}
    Catalogued Galaxies and Quasars in the IRAS Survey 1985, 
    prepared by Lonsdale, C.J., et al. (JPL)
\bibitem[Combes(1991)]{C91}
    Combes, F. 1991, \araa, 29, 195
\bibitem[Condon(1992)]{C92}
    Condon, J.J. 1992, \araa, 30, 575
\bibitem[Crosthwaite et al.(2001)]{CTHLMH01}
    Crosthwaite, L.P., et al. 2001, \aj, 122, 797
\bibitem[Crosthwaite et al.(2002)]{CTBHM02}
    Crosthwaite, L.P., et al. 2002, \aj, 123, 1892
\bibitem[Dahmen et al.(1998)]{DHWM98}
    Dahmen, G., et al.  1998, \aap, 331, 959
\bibitem[DeGioia-Eastwood et al.(1984)]{DGSS84}
    DeGioia-Eastwood, K., et al.  1984, \apj, 278, 564
\bibitem[de Vaucouleurs, et al.(1976)]{dVdVC76}
    de Vaucouleurs, G., de Vaucouleurs, A. \& Corwin, H. 1976,
    {\em Second Reference Catalogue of Bright Galaxies},
    Univ. of Texas Press
\bibitem[Dickman et al.(1986)]{DSS86}
    Dickman, R.L., Snell, R.L. \& Schloerb, F.P. 1986, \apj, 309, 326
\bibitem[Eastman et al.(1996)]{ESK96}
    Eastman, R.G., et al. 1996, \apj, 466, 911
\bibitem[Elmegreen(1989)]{E89}
    Elmegreen, B.G. 1989, \apj, 338, 178
\bibitem[Elmegreen(1993)]{E93}
    Elmegreen, B.G. 1993, \apj, 411, 170
\bibitem[Elmegreen(2002)]{E02}
    Elmegreen, B.G. 2002, \apj, 577, 206
\bibitem[Elmegreen \& Parravano(1994)]{EP94}
    Elmegreen, B.G. \& Parravano, A.  1994, \apj, 435, L121
\bibitem[Engargiola(1991)]{E91}
    Engargiola, G.  1991, \apjs, 76, 875
\bibitem[Ferguson et al.(1998)]{FWGH98}
    Ferguson, A.M.N., et al.  1998, \apj, 506, L19
\bibitem[Gordon et al.(1968)]{GRR68}
    Gordon, K.J., Remage, N.H. \& Roberts, M.S.  1968, \apj, 154, 845
\bibitem[Helou \& Bicay(1993)]{HB93}
    Helou, G. \& Bicay, M.D. 1993, \apj, 415, 93
\bibitem[Hodge \& Kennicutt(1983)]{HK83}
    Hodge, P.W. \& Kennicutt, R.C.  1983, \aj, 88, 296
\bibitem[Honma et al.(1995)]{HSA95}
    Honma, M., Sofue, Y. \& Arimoto, N. 1995, \aap, 304, 1
\bibitem[Hunter et al.(1997)]{Hea97}
    Hunter, S.D., et al. 1997, \apj, 481, 205
\bibitem[Ishizuki et al.(1990)]{Iea90}
    Ishizuki, S. et al.  1990, \apj, 355, 436
\bibitem[Kamphuis \& Sancisi(1993)]{KS93}
    Kamphuis, J. \& Sancisi, R.  1993, \aap, 273, L31
\bibitem[Karachentsev et al.(2000)]{KSH00}
    Karachentsev, I.D., Sharina, M.E. \& Hutchmeier, W.K.  
    2000, \aap, 362, 544
\bibitem[Kennicutt(1998a)]{K98a}
    Kennicutt, R.C. 1998a, \apj, 498, 541
\bibitem[Kennicutt(1998b)]{K98b}
    Kennicutt, R.C. 1998b, \araa, 36, 189
\bibitem[Kulkarni \& Heiles(1987)]{KH87}
    Kulkarni, S.R., \& Heiles, C. 1987,
    in {\em Interstellar Processes}, ed. Hollenbach \& Thronson, p. 87
\bibitem[Kutner \& Ulich(1981)]{KU81}
    Kutner, M.L. \& Ulich, B.L.  1981, \apj, 250, 341
\bibitem[Lacey et al.(1997)]{LDG97}
    Lacey, C., Duric, N. \& Goss, W.M.  1997, \apjs, 109, 417
\bibitem[Lequeux et al.(1993)]{LAG93}
    Lequeux, J., Allen, R.J. \& Guilloteau, S. 1993, \aap, 280, L23
\bibitem[Leroy et al.(2005)]{LBSB05}
    Leroy,A. et al. 2005, \apj, 625, 763
\bibitem[Madden et al.(1993)]{Mea93}
    Madden, S.C. et al. 1993, \apj, 407, 579
\bibitem[Maloney \& Black(1988)]{MB88}
    Maloney, P. \& Black, J.H. 1988, \apj, 325, 389
\bibitem[Mangum(1996a)]{M96a}
    Mangum, J.G. 1996, 
    {\em User's Manual for the NRAO 12m Millimeter-Wave Telescope},
    NRAO 12m Publication
\bibitem[Mangum(1996b)]{M96b}
    Mangum, J.G. 1996, 
    {\em On The Fly Observing at the 12m},
    NRAO 12m Publication
\bibitem[Mangum(1997)]{M97}
    Mangum, J.G. 1997,
    {\em Equipment and Calibration Status for the NRAO 12m Telescope},
    NRAO 12m Publication
\bibitem[Mason \& Wilson (2004)]{MW04}
    Mason, A. M., \& Wilson, C. D. 2004, \apj, 612, 860
\bibitem[Mead \& Kutner(1988)]{MK88}
    Mead, K.N. \& Kutner, M.L. 1988, \apj, 330, 399
\bibitem[Meier \& Turner(2004)]{MT04}
    Meier, D.S. \& Turner, J.L. 2004, \aj, 127, 2069
\bibitem[Meier \& Turner(2001)]{MT01}
    Meier, D.S. \& Turner, J.L. 2001, \apj, 551, 687
\bibitem[Meier, Turner, \& Beck (2002)]{MTB02}
    Meier, D.S., Turner, J.L., \& Beck, S. C. 2002, \aj, 124, 877
\bibitem[Morris \& Lo(1978)]{ML78}
    Morris, M. \& Lo, K.Y.  1978, \apj, 223, 803
\bibitem[Murgia et al.(2002)]{MCMG02}
    Murgia, M., et al.  2002, \aap, 385, 412
\bibitem[Nikiforov et al.(2000)]{NPN00}
    Nikiforov, I.I., Petrovskaya, I.V. \& Ninkovic, S.  2000, 
    {\em Small Galaxy Groups, ASP Conference Serier}, 209, 399
\bibitem[Oka et al.(1996)]{OHHHS96}
    Oka, T., et al.  1996, \apj, 460, 334
\bibitem[Oka et al.(1998)]{OHHHS98}
    Oka, T., et al.  1998, \apj, 493, 730
\bibitem[Pisano \& Wilcots(2000)]{PW00}
    Pisano, D.J. \& Wilcots, E.M. 2000, \mnras, 319, 821
\bibitem[Regan \& Vogel(1995)]{RV95}
    Regan, M.W. \& Vogel, S.N. 1995, \apj, 452, L21
\bibitem[Regan et al.(2001)]{Rea01}
    Regan, M.W. et al. 2001, \apj, 561, 218
\bibitem[Rogstad et al.(1973)]{RSR73}
    Rogstad, D.H., Shostak, G.S. \& Rots, A.H.  1973, \aap, 22, 111
\bibitem[Rownd \& Young(1999)]{RY99}
    Rownd, B.K. \& Young, J.S. 1999, \aj, 118, 670
\bibitem[Saha et al.(2002)]{Sea02}
    Saha, A, Claver, J. \& Hoessel, J.G. 2002, \aj, 124, 839
\bibitem[Sakamoto(1996)]{S96}
    Sakamoto, S. 1996, \apj, 462, 215
\bibitem[Sakamoto et al.(1995)]{SHHHO95}
    Sakamoto, S., et al.  1995, \apjs, 100, 125
\bibitem[Sakamoto et al.(1997)]{SHHHO97}
    Sakamoto, S., et al.  1997, \apj, 486, 290
\bibitem[Sakamoto et al.(1999)]{SOIS99}
    Sakamoto, S., et al.  1999, \apjs, 124, 403
\bibitem[Sauvage \& Thuan(1992)]{ST92}
    Sauvage, M. \& Thuan, T.X. 1992, \apj, 396, L69
\bibitem[Schlegel(1994)]{S94}
    Schlegel, E. M. 1994, \apj, 434, 523
\bibitem[Schmidt(1959)]{S59}
    Schmidt, M. 1959, \apj, 129, 243
\bibitem[Scoville \& Sanders(1987)]{SS87}
    Scoville, N. Z. \& Sanders, D.B. 
    1987, in {\em Interstellar Processes}, ed. Hollenbach \& Thronson, p 21
\bibitem[Scoville et al. (1987)]{SY87} 
    Scoville, N. Z., et al.  1987, \apjs, 63, 821
\bibitem[Sharina et al.(1997)]{SKT97}
    Sharina, M.E., Karachentsev, I.D. \& Tikhonov, N.A. 1997, AstL, 23, 373
\bibitem[Sofue et al.(1988)]{SDINH88}
    Sofue, Y.,  et al. 1988, \pasj, 40, 511
\bibitem[Solomon et al. (1987)]{SRBY87}
    Solomon, P.M., et al. 1987, \apj, 319, 730
\bibitem[Spaans et al. (1994)]{Sea94}
    Spaans, M., et al. 1994, \apj, 437, 270
\bibitem[Strong et al.(1988)]{Sea88}
    Strong, A.W. et al.  1988, \aap, 207, 1
\bibitem[Suchkov et al.(1993)]{SAH93}
    Suchkov, A., Allen, R.J. \& Heckman, T.M. 1993, \apj, 412, 542
\bibitem[Tacconi \& Young(1986)]{TY86}
    Tacconi, L.J. \& Young, J.S.  1986, \apj, 308, 600
\bibitem[Tacconi \& Young(1989)]{TY89}
    Tacconi, L.J. \& Young, J.S.  1989, \apjs, 71, 455
\bibitem[Tacconi \& Young(1990)]{TY90}
    Tacconi, L.J. \& Young, J.S.  1990, \apj, 352, 595
\bibitem[Tilanus \& Allen(1989)]{TA89}
    Tilanus, R.P.J. \& Allen, R.J.  1989, \apj, 339, L57
\bibitem[Tilanus \& Allen(1993)]{TA93}
    Tilanus, R.P.J. \& Allen, R.J.  1993, \aap, 274, 707
\bibitem[Tuffs et al.(1996)]{Tea96}
    Tuffs, R.J. et al.  1996, \aap, 315, L149
\bibitem[Ulich \& Haas(1976)]{UH76}
    Ulich, B.L. \& Hass, R.W.  1976, \apjs, 30, 247
\bibitem[van der Kruit et al.(1977)]{KAR77}
    van der Kruit, P.C., Allen, R.J. \& Rots, A.H.  1977, \aap, 55, 421
\bibitem[Wall et al.(1993)]{Wea93}
    Wall, W.F., et al.  1993, \apj, 414, 98
\bibitem[Walsh et al.(2002)]{Wea02}
    Walsh, W., et al.  2002, \aap, 388, 7
\bibitem[Weliachew et al.(1988)]{WCC88}
    Weliachew, L., Casoli, F. \& Combes, F. 1988, \aap, 199, 29
\bibitem[Wiklind et al.(1990)]{WRHB90}
    Wiklind, T., et al.  1990, \aap, 232, L11
\bibitem[Wilson(1991)]{W95}
    Wilson, C.D. 1995, \apj, 448, L97
\bibitem[Wilson \& Scoville(1991)]{WS91}
    Wilson, C.D. \& Scoville, N. 1991, \apj, 370, 184
\bibitem[Wong \& Blitz(2002)]{WB02}
    Wong, T. \& Blitz, L. 2002, \apj, 569, 157
\bibitem[Wouterloot et al.(1988)]{WBH88}
    Wouterloot, J.G.A., Brand, J. \& Henkel, C. 1988, \aap, 191, 323
\bibitem[Young \& Scoville(1982)]{YS82}
    Young, J.S. \& Scoville, N.Z.  1982, \apj, 258, 467
\bibitem[Young \& Scoville(1991)]{YS91}
    Young, J.S. \& Scoville, N.Z.  1991, \araa, 29, 581
\bibitem[Young et al.(1995)]{Yea95}
    Young, J.S., et al.  1995, \apjs, 98, 219
\end{thebibliography}
\end{document}